\documentclass[11pt,reqno,preprint]{article}
\usepackage{jheppub} % for details on the use of the package, please see the JINST-author-manual

\usepackage{enumitem}
\usepackage{shuffle}
\usepackage{float}
\usepackage{tikz-feynman}
\usepackage{comment}

\usetikzlibrary{decorations.pathreplacing}
\usetikzlibrary{calc,tikzmark}

\newcommand{\CP}{\mathbb{CP}}
\newcommand{\PT}{\mathbb{PT}}
\newcommand{\PTs}{\mathbb{PT}^{3|4}}
\newcommand{\cO}{\mathcal{O}}
\newcommand{\cA}{\mathcal{A}}
\newcommand{\cZ}{\mathcal{Z}}
\newcommand{\cN}{\mathcal{N}}

\newcommand{\SL}{\mathrm{SL}}

\newcommand{\bk}[1]{\left\langle #1 \right\rangle} 
\newcommand{\spa}[2]{\left\langle#1#2\right\rangle}
\newcommand{\spb}[2]{\left[#1#2\right]}
\newcommand{\la}{\left\langle}
\newcommand{\ra}{\right\rangle}
\newcommand{\bthree}{\tfrac{\sqrt{2}}{3!}\tr(B^3)}
\newcommand{\jt}{\tilde{J}}

\newcommand{\lamt}{\tilde{\lambda}}
\newcommand{\F}{\mathcal{F}}
\newcommand{\sig}{\sigma}
\newcommand{\trg}[1]{\mathrm{tr}_\mathfrak{g}\left(#1\right)}
\newcommand{\spdenom}[1]{\spa{1}{2}\spa{2}{3}\cdots\spa{#1}{1}}
\newcommand{\spdenomsig}[1]{\spa{\sig_1}{\sig_2}\spa{\sig_2}{\sig_3}\cdots\spa{\sig_{#1}}{\sig_1}}
\newcommand{\tr}{\mathrm{tr}}
\newcommand{\cJ}{\mathcal{J}}

\title{\boldmath Form factors of $\mathcal{N}=4$ self-dual Yang-Mills from the chiral algebra bootstrap}

\author[1]{Jaazib Charanya}
\author[1]{Anthony Morales}
\author[1]{and Natalie M. Paquette}
\affiliation[1]{University of Washington, Department of Physics \\ 3910 15th Ave NE, Seattle, WA 98195, U.S.A}

\emailAdd{jazchar@uw.edu}
\emailAdd{moralesa@uw.edu}
\emailAdd{npaquett@uw.edu}

\abstract{
The chiral algebra bootstrap (CAB) is a novel bootstrap program for form factors in quantum-integrable self-dual gauge theories, some of which in turn are helicity amplitudes in the corresponding gauge theories. The singularities that recursively generate a given (loop-level) form factor are holomorphic collinear splitting functions, equivalently celestial chiral algebra OPEs,  of the self-dual theory. In this note, we apply the chiral algebra bootstrap to the simple example of self-dual 4d $\mathcal{N}=4$ super Yang-Mills (SDSYM). We use a combination of twistor space input, Koszul duality, supersymmetry, and associativity to obtain the all-loop holomorphic collinear splitting functions for SDSYM. We also use associativity to provide a simple proof of the conjecture that there are no double-poles in the loop-level OPEs for this theory. We conclude by computing several form factors, including both a reproduction of several known results and novel form factors up to two loops involving insertions of powers of the anti-self-dual field strength. These form factors compute a supersymmetric version of Higgs amplitudes in the self-dual sector. Detailed sample computations are provided to familiarize the reader with the CAB method.
}

\begin{document}
\maketitle
\flushbottom

\section{Introduction \& Background}
In recent years, a proliferation of techniques have been developed for perturbatively computing scattering amplitudes and form factors in planar $\mathcal{N}=4$ supersymmetric Yang-Mills (SYM) theory. Chief among these techniques are bootstrap methods, which utilize boundary data supplied by the flux-tube representation or the pentagon operator product expansion (OPE) \cite{Basso:2013vsa,Basso:2013aha,Basso:2014koa,Basso:2014nra,Basso:2015rta,Basso:2015uxa} and its extension to form factors of protected operators \cite{Sever:2020jjx,Sever:2021nsq,Sever:2021xga,Basso:2023bwv}. These techniques fix the analytic form of observables while avoiding Feynman diagrams. Bootstrap methods have been used to compute observables to high multiplicity and high loop orders \cite{Dixon:2011nj,Dixon:2011pw,Dixon:2013eka,Dixon:2014iba,Dixon:2014voa,Dixon:2015iva,Dixon:2016apl,Dixon:2016nkn,Dixon:2020bbt,Dixon:2022rse,Dixon:2023kop,Dixon:2024yvq,Drummond:2014ffa,Drummond:2018caf,Caron-Huot:2016owq,Caron-Huot:2019vjl,Basso:2024hlx,Henn:2024pki}. This plethora of data has led to important discoveries in planar $\mathcal{N}=4$ SYM, such as the amplitude/Wilson-loop duality \cite{Alday:2007hr,Drummond:2007aua,Brandhuber:2007yx,Drummond:2007au,Alday:2008yw,Adamo:2011pv,Brandhuber:2010ad,Alday:2007he,Bern:2008ap,Drummond:2008aq,Maldacena:2010kp,Ben-Israel:2018ckc}, dual conformal invariance \cite{Alday:2007hr,Drummond:2006rz,Bern:2007ct,Drummond:2008vq}, and the still-mysterious antipodal duality \cite{Dixon:2021tdw,Dixon:2022xqh,Dixon:2025zwj}.

Less progress has been made in the non-planar regime of $\mathcal{N}=4$ SYM. There have been successes in computing Sudakov form factors of the stress-tensor multiplet, with results up to five loops \cite{Yang:2016ear,Boels:2017ftb,Lee:2021lkc}, while 3 point form factors of the stress-tensor multiplet have been computed up to three loops \cite{Lin:2021qol,Guan:2023gsz,Guan:2025awp}. In this note, we will demonstrate the utility of a complementary bootstrap called the chiral algebra bootstrap, which computes certain full-color, non-planar, all-multiplicity form factors in self-dual $\mathcal{N}=4$ SYM. 

The chiral algebra bootstrap (CAB) \cite{CP} provides a perturbative bridge between, on the one hand, observables in special four-dimensional gauge theories and, on the other hand, two–dimensional chiral algebra correlators. It has been used recently to recover $n$-point helicity amplitudes at one loop (with one negative-helicity gluon) \cite{CPassoc} and two loops (with all positive-helicity gluons) \cite{CosQCD} in, for example, $G=SU(3)$ gauge theory coupled to $N_f = 9$ massless fundamental fermions. Remarkably, by cleverly combining known CAB results for such special gauge theories with additional tree-level and one-loop amplitudes, one can even obtain new two loop $n$-point all-plus amplitudes for massless $G= SU(3)$ QCD with arbitrary $N_f$ \cite{anthony}.  

The gauge theories to which the CAB applies are those that can be uplifted to local, holomorphic gauge theories on twistor space; these are self-dual gauge theories coupled to massless matter that guarantees  that the quantum theory on twistor space is free of gauge anomalies \cite{CosAnomaly}. The cancellation of the twistor space anomaly is equivalent to quantum integrability of the corresponding four dimensional theory, which manifests as the existence of infinitely many currents whose OPEs define a chiral algebra. This chiral algebra may be thought of as being supported on the ``celestial sphere,'' which is geometrized as the twistor space $\mathbb{CP}^1$. For details on the origin of this chiral algebra, see ref.~\cite{CP}. In this paper, we will construct the chiral algebra associated to SDSYM by implementing the following procedure \cite{FP}: (i) write the most general meromorphic OPEs compatible with kinematics and symmetries (in this paper, including supersymmetry); and (ii) associativity of the OPE (equivalently, at the level of the Lie algebra modes, Jacobi identities). It turns out, perhaps surprisingly, that these are enough to fix the OPE data at all loop orders. In particular, the symmetries of the local twistorial uplift are highly constraining, and they allow for only low order poles in the collinear limit, even at arbitrary loop number.

With the chiral algebra in hand, we compute several all-multiplicity form factors up to two loops for operators $\textrm{tr}(B^m)$, $m=2, 3$, along with the supersymmetric completion of the former, which coincide on-shell with insertions of powers of the anti-self-dual field strength. Using the all-loop chiral algebra, this program can be extended to higher loops and other $m$, but may need to be supplemented, as we explain in the text, with small numbers of Feynman diagram computations. Although our purpose in this note is to illustrate applications of the CAB in non-planar computations, we verify our answers with known results, where available, and supersymmetric Ward identities along the way. We also view this note as a helpful warm-up for tackling higher-loop $\textrm{tr}(B^m)$ form factors in non-supersymmetric gauge theories, which we plan to report on in future work.

\subsection{Plan of paper}  Our plan for the rest of the paper is as follows. We conclude this introduction with a brief summary of conventions and notation used in the sequel. We review $\mathcal{N}=4$ self-dual SYM in its twistor space and spacetime formulations in \S \ref{sec:hCS}. We present its all-loop holomorphic collinear splitting functions, encoded in chiral algebra OPEs, in \S \ref{sec:all-loop} and provide an alternative, elementary proof that this algebra is free from any second order poles in \S \ref{sec:doublepoles}. We compute $\textrm{tr}(B^2)$ and related form factors up two loops, which agree with known results for SYM amplitudes in the same helicity sectors, in \S \ref{sec:b2formfactors}. In \S \ref{sec:bcubed}, we compute new $\textrm{tr}(B^3)$ form factors, up to two loops. The reader who wishes to go directly to the formulas can consult eqs.~\eqref{F3}-\eqref{F0}. Detailed formulas for the chiral algebra OPEs can be found in appendices \ref{sec:appcoeffs}, \ref{app:recursion}, and the proof of a supplemental result can be found in appendix \ref{app:proof}.

\subsection{Conventions} We briefly establish the conventions we use in the sequel and summarize some requisite elementary facts about twistor space. We work in four (complexified) spacetime dimensions $x^{\alpha\dot\alpha}$ with $\alpha,\dot\alpha=1,2$. Null momenta are given by $p_{\alpha\dot\alpha}=\lambda_\alpha \tilde\lambda_{\dot\alpha}$, so the rank-one condition $p^2=0$ is automatic and collinearity corresponds to alignment of spinors. We work on both projective twistor space $\PT$ and its $\mathcal{N}=4$ supersymmetric extension $\PTs$. 

We give (bosonic) twistor space $\mathbb{PT}$ the following coordinates, subject to projective equivalence:
\begin{equation}
	Z^I=(\lambda_\alpha,\mu^{\dot\alpha})\,,\qquad Z\sim t\,Z\quad (t\in\mathbb{C}^\times).
\end{equation} \\
The incidence relation
\begin{equation}
	\mu^{\dot\alpha}=x^{\alpha\dot\alpha}\lambda_\alpha
	\label{eq:incidence}
\end{equation}
identifies each spacetime point $x^{\alpha\dot\alpha}$ with a holomorphic line $L_x\simeq \CP^1\subset \PT$. Conversely, a point $Z\in \PT$ corresponds to a self-dual totally null two-plane in spacetime, as the locus of $x$ obeying eq.~\eqref{eq:incidence} is an affine two-plane with self-dual tangent. Two spacetime points $x,y$ are null separated iff the lines $L_x,L_y$ intersect at a single twistor point.

The complexified conformal group $\SL(4,\mathbb{C})$ acts linearly on $Z^I$, and the Euler vector $E=Z^I \partial/\partial Z^I$ generates projective rescalings $Z\mapsto tZ$. A twistor function $f(Z)$ has homogeneity $w$ if $Ef=w f$. This $w$ is mapped to 4d helicity via the Penrose transform, and, hence, it serves as a simple selection rule for OPE poles and loop counterterms.

The Penrose transform is the statement that massless fields of helicity $h$ correspond to Dolbeault cohomology classes on $\PT$:
\begin{equation}
	\quad \text{helicity } h \;\longleftrightarrow\; H^{(0,1)}\!\left(\PT,\cO(2h-2)\right)\quad
	\label{eq:penrose}
\end{equation}
where $\cO(k)$ is the degree-$k$ line bundle on $\CP^3$, constructed in practice using twistor functions of the corresponding homogeneity.

We similarly give supertwistor space $\PTs$ the coordinates
\begin{equation}
	\cZ^A=(Z^I\,|\,\eta^a)=(\lambda_\alpha,\mu^{\dot\alpha}\,|\,\eta_a),\qquad a=1,\dots,4,
\end{equation}
and supersymmetric incidence relations
\begin{equation}
	\mu^{\dot\alpha}=x^{\alpha\dot\alpha}\lambda_\alpha,\qquad
	\eta_a=w_a^{\alpha}\lambda_\alpha .
\end{equation}
$w_a^{\alpha}$ is a fermionic coordinate on chiral superspace. Projective rescaling $\cZ\sim t\,\cZ$ acts with weight $+1$ on $\lambda,\mu,\eta$. Holomorphic densities live naturally in the Berezinian line; the holomorphic volume form $\Omega^{3|4}$ has weight 0 and is globally defined.

We switch to affine coordinates by rescaling coordinates by a factor of $\frac1{\lambda_1}$. This sets one of the projective coordinates to unity and scales the rest.  
\begin{align}
	z:=\frac{\lambda_2}{\lambda_1},\quad
	v^{\dot{1}}:=\frac{\mu^{\dot 1}}{\lambda_1},\quad
	v^{\dot{2}}:=\frac{\mu^{\dot 2}}{\lambda_1},
	\quad \theta_a:=\frac{\eta_a}{\lambda_1}
	\label{eq:affine}
\end{align}
In what follows, we will make use of both affine coordinates $(z,v^{\dot{1}},v^{\dot{2}};\theta_a)$ and projective coordinates $(\lambda_1,\lambda_2,\mu^{\dot{1}},\mu^{\dot{2}};\eta_a)$.

\section{SDSYM on twistor space}
\label{sec:hCS}

SDSYM has a well-known action on the $\mathcal{N}=4$ super-twistor space $\mathbb{PT}^{3|4}$, which is an open subset of $\mathbb{CP}^{3|4}$ \cite{WittenTwistor, Boels:2006ir, Mason}, on which it takes the form of a holomorphic Chern-Simons theory whose fundamental field is a superconnection.

Explicitly, let $\cA=\cA_{\bar I}(\cZ,\bar\cZ)\, d\bar Z^{\bar I}$ be a Lie-algebra valued $(0,1)$ superconnection on $\PTs$ of projective weight zero, with coordinates $\cZ=(Z\,|\,\eta)=(\lambda_\alpha,\mu^{\dot\alpha}\,|\,\eta_a)$. The holomorphic Chern--Simons (hCS) functional is
\begin{equation}
    \label{eq:hCS}
	S_{\mathrm{hCS}}[\cA]
	\;=\frac{i}{2\pi}\int_{\PTs}\!\Omega^{3|4}\;
	\tr\!\left(\frac{1}{2}\cA\wedge\bar\partial\cA+\frac{1}{3}\,\cA\wedge\cA\wedge\cA\right),
\end{equation}
where $\Omega^{3|4}$ is the holomorphic Berezinian on $\PTs$ of total projective weight zero. The Euler--Lagrange equation is the holomorphic flatness condition
\begin{equation}
	\bar\partial\cA + \cA\wedge \cA \;=\; 0,
	\label{eq:hCS-EOM}
\end{equation}
which reproduces the self-dual Yang--Mills equations $F_-(A) = 0$ on spacetime.

We expand $\cA$ in the fermionic coordinates:
\begin{equation}
	\cA(Z,\eta)
	\;=\;
	a
	\;+\; \eta_i \lambda^i
	\;+\; \frac{1}{2} \eta_i\eta_j \phi^{ij}
	\;+\; \frac{1}{3!}\epsilon^{ijkl}\eta_i\eta_j\eta_k \tilde\lambda_{l}
	\;+\; \frac{1}{4!}\epsilon^{ijkl}\eta_i\eta_j\eta_k\eta_l\, b,
	\label{eq:super-connection-expansion}
\end{equation}
where $a,\lambda,\phi,\tilde\lambda,g$ are $\PT$-valued $(0,1)$-forms of definite homogeneity in $Z$ (respectively: weights 0, -1, -2, -3, -4), valued in $\mathfrak g$. Upon dimensional reduction along the twistor spheres (equivalently, using the Penrose-Ward correspondence), the component fields furnish the on-shell $\cN=4$ multiplet restricted to the self-dual sector. In particular, $a$ reduces to the positive-helicity gluons, $\lambda^i$ are four positive-helicity gauginos, $\phi^{ij} = -\phi^{ji}$ six real scalars which transform in the \textbf{6} of the $SU(4)_R$-symmetry, $\tilde{\lambda}_l$ the four negative-helicity gauginos, and $b$ the negative-helicity gluons.  \\

The hCS action is invariant under the gauge transformation
\begin{equation}
	\delta_\varepsilon \cA \;=\; \bar\partial \varepsilon + [\cA,\varepsilon],
	\qquad
	\varepsilon
	= \alpha + \eta_i\beta^i + \frac{1}{2} \eta_i\eta_j\gamma^{ij}
	+ \frac{1}{3!}\epsilon^{ijkl}\eta_i\eta_j\eta_k \tilde\beta_{l}
	+ \frac{1}{4!}\epsilon^{ijkl}\eta_i\eta_j\eta_k\eta_l \kappa,
	\label{eq:supergauge}
\end{equation}
with $\alpha,\beta,\gamma,\tilde{\beta},\kappa$ the gauge parameters given by $(0,0)$-form fields of the appropriate projective weight.

\subsection{Actions on $\PT$ and spacetime}

It will be convenient for some of our subsequent computations to perform the Grassmann integration in eq.~\eqref{eq:hCS} over $d^4 \eta$ to obtain the component action on $\mathbb{PT} \subset \mathbb{CP}^3$. The result is
\begin{equation}
	S_{\PT=}
	\int_{\PT}\,\Omega^{3,0}\wedge
	\tr\!\Big(b\wedge F_{0,2}(a)+\tilde{\lambda}_i\wedge\bar{D}\lambda^i+\tfrac{1}{2}\epsilon_{ijkl}\,\phi^{ij}\wedge\bar{D}\phi^{kl}+\tfrac{1}{2}\epsilon_{ijkl}\,\lambda^i\wedge\lambda^j\wedge\phi^{kl}\Big),
	\label{eq:twistor-action}
\end{equation}
with the covariant derivative and field strength given by
\begin{align}
	\bar{D}:=\bar{\partial}+[a,\cdot], \qquad F_{0,2}(a):=\bar{\partial}a+a\wedge a.
\end{align}
This twistor space action enjoys the following gauge transformations obtained from eq.~\eqref{eq:supergauge}:
\begin{align}
	\delta a \;&=\; \bar\partial \alpha + [a,\alpha],
	\label{eq:delta-a}
	\\
	\delta \lambda^i \;&=\; \bar\partial \beta^i + [a,\beta^i] + [\lambda^i,\alpha],
	\label{eq:delta-psi}
	\\
	\delta \phi^{ij} \;&=\; \bar\partial \gamma^{ij} + [a,\gamma^{ij}] - \{\lambda^{i},\beta^{j}\} +\{\lambda^j,\beta^i\}+ [\phi^{ij},\alpha],
	\label{eq:delta-phi}
	\\
	\delta \tilde\lambda_{i} \;&=\; \bar\partial \tilde\beta_{i} + [a,\tilde\beta_{i}] - \tfrac{\epsilon_{ijkl}}{2}\big( [\lambda^{j},\gamma^{kl}] + [\phi^{jk},\beta^{l}]\big)+[\tilde\lambda_{i},\alpha],
	\label{eq:delta-psitilde}
	\\
	\delta b \;&=\; \bar\partial \kappa + [a,\kappa] 
	+ \{\lambda^{i},\tilde\beta_{i}\} + \tfrac{\epsilon_{ijkl}}{4}[\phi^{ij},\gamma^{kl}] - \{\tilde\lambda_{i},\beta^{i}\}+ [b,\alpha],
	\label{eq:delta-g}
\end{align}
with $\gamma^{ij}:=\frac12\epsilon^{ijkl}\gamma_{kl}$ and $\phi^{ij}:=\frac12\epsilon^{ijkl}\phi_{kl}$. For later convenience, we expand these transformations in a Lie-algebra basis $X=X^at_a$, $[t_a,t_b]=f_{ab}^ct_c$:
\begin{align}
	\delta a^{a} \;&=\; \bar\partial \alpha^{a} + f_{bc}^a\,a^{b}\alpha^{c},
	\\
	\delta \lambda^{ai} \;&=\; \bar\partial \beta^{ai} + f_{bc}^a\!\left(a^{b}\beta^{ci} + \lambda^{bi}\alpha^{c}\right),
	\\
	\delta \phi^{aij} \;&=\; \bar\partial \gamma^{aij}
	+ f_{bc}^a\!\left(a^{b}\gamma^{cij} -\lambda^{b[i}\beta^{cj]} + \phi^{bij}\alpha^{c}\right),
	\\
	\delta \tilde\lambda^{a}_i \;&=\; \bar\partial \tilde\beta^{a}_i
	+ f_{bc}^a\!\left(a^{b}\tilde\beta^{c}_i  - \tfrac{\epsilon_{ijkl}}{2}(\lambda^{bj}\gamma^{ckl}+\phi^{bjk}\beta^{cl})+ \tilde\lambda^{b}_i\alpha^{c}\right),
	\\
	\delta b^{a} \;&=\; \bar\partial \kappa^{a}
	+ f_{bc}^a\!\left(a^{b}\kappa^{c} 
	+ \lambda^{bi}\tilde\beta^{c}_i
	+ \tfrac{\epsilon_{ijkl}}{4}\,\phi^{bij}\gamma^{ckl}
	- \tilde\lambda_{i}^b\beta^{ci}+ b^{b}\alpha^{c}\right).
\end{align}
We can package the gauge transformations nicely in the BRST-BV formalism by promoting the gauge parameters to a ghost superfield
\begin{equation}
	C \;=\; \alpha + \eta_i\beta^i + \tfrac12 \eta_i\eta_j\gamma^{ij}
	+ \tfrac{1}{3!}\epsilon^{ijkl}\eta_i\eta_j\eta_k \tilde\beta_{l}
	+ \tfrac{1}{4!}\epsilon^{ijkl}\eta_i\eta_j\eta_k\eta_l \kappa,
\end{equation}
with Grassmann-odd components. The BRST operator acts as
\begin{equation}
	s\,\cA \;=\; \bar\partial C + [\cA,C],
	\qquad
	s\,C \;=\; -\tfrac{1}{2}[C,C].
	\label{eq:BRSTsuper}
\end{equation}

Finally, we introduce antifields $\cA^*$ and $C^*$ of complementary form degree so that the BV action reads
\begin{equation}
	S_{\rm BV} \;=\; S_{\PT}  + \int_{\PT}\Omega^{3,0}\wedge
	\tr\!\left(
	a^* \wedge s a
	+ \lambda^{*}_{a} \wedge s\lambda^a
	+ \tfrac12 \phi^{*}_{ab} \wedge s\phi^{ab}
	+ \tilde\lambda^{*a} \wedge s\tilde\lambda_{a}
	+ b^* \wedge s b
	+ C^*\wedge sC
	\right).
	\label{eq:BVminimal}
\end{equation}

Because this holomorphic field theory does not suffer from any gauge anomalies on twistor space \cite{CosAnomaly, CP}, the theorem of ref.~\cite{CP} guarantees that a well-defined, associative chiral algebra may be associated to the 4d SDSYM theory \eqref{eq:sdN4}, whose OPEs encode all its collinear splitting functions. Even in SDSYM, there are nontrivial, non-factorizing contributions to collinear splitting functions at each loop order when working with complexified momenta; these contribute nontrivially to form factors.

Being guaranteed the existence of this chiral algebra (equivalently, quantum integrability of the 4d theory), we may use the techniques of Koszul duality \cite{CPsugra, CP, PW}, associativity \cite{CPassoc, roland, FP}, and covariance under supersymmetry transformations \cite{Tropper} to obtain the OPEs. Indeed, we need only perform a slight modification of the techniques of ref.~\cite{FP}, where this program was carried out for non-supersymmetric gauge theories. We will see that the result in SDSYM is much simpler.

After the standard gauge-fixing and dimensional reduction from eq.~\eqref{eq:twistor-action}, we recover the usual SDSYM action on spacetime. We denote the 4d fields by capital letters for clarity: 
\begin{equation}
	S_\mathrm{SDSYM}[\cA]=\int d^4x\;{\rm tr}\left\{{B}_{ab} {F}^{-, ab}
	+i\tilde\Lambda_{i}^aD_{a\dot{a}}\Lambda^{i\,\dot{a}} +
	\tfrac{1}{2}\epsilon_{ijkl} D^a_{\;\dot{a}}\Phi^{ij}D_a^{\;\dot{a}}\Phi^{kl}+
	\tfrac{1}{2}\epsilon_{ijkl}\Phi^{ij}\Lambda^{k}_{\dot{a}}\Lambda^{l\,\dot{a}}\right\},
	\label{eq:sdN4}
\end{equation}
where ${F}^-_{ab}$ denotes the anti-self-dual part of the field strength, ${B^{ab}}$ is an anti-self-dual two-form imposing the self-duality equation of motion, $\Lambda$ and $\tilde\Lambda$ are Weyl fermions, $\Phi^{ab}=-\Phi^{ba}$ are the six real scalars packed into an $SU(4)$-antisymmetric tensor with $\Phi^{ij}=\frac12\varepsilon^{ijkl}\Phi_{kl}$, and $D^{a\dot{a}}$ is the usual spacetime covariant derivative.

\section{All-loop holomorphic collinear splitting functions}\label{sec:all-loop}

In preparation for computing form factors, we adapt the techniques of ref.~\cite{FP} to the case of SDSYM, to which we refer the reader for more details. In ref.~\cite{FP}, all twistorial gauge theories of BF-type on $\mathbb{PT}$, coupled to fermionic matter and/or Green-Schwarz-type scalars, were treated uniformly, and their all-loop chiral algebra OPEs were determined. Here, from the perspective of the action \eqref{eq:twistor-action}, we need only account for the addition of ordinary adjoint-valued scalar degrees of freedom and the Yukawa coupling. 

The chiral algebra generators correspond to massless momentum eigenstates $\mathcal{C}[\lamt](z)$ with complexified momenta $p_{a\dot{a}}=\lambda_a\lamt_{\dot{a}}$ and $\lambda_a=(1,z)$. Notice that the two factors of the complexified Lorentz algebra $\mathfrak{sl}(2, \mathbb{C})_- \times \mathfrak{sl}(2, \mathbb{C})_+$ have rather asymmetric roles on twistor space. The former acts on the chiral algebra as a 2d holomorphic conformal symmetry, whereas the latter, which rotates the twistor fibre coordinates $v^{\dot{1}}, v^{\dot{2}}$ (see eq.~\eqref{eq:affine}), acts as a flavor symmetry. It is convenient to Taylor expand each momentum eigenstate $\mathcal{C}[\tilde{\lambda}](z)$ in powers of energy $\omega$, as follows:
\begin{equation}
	\label{eq:softmodes}
	\mathcal{C}[\tilde{\lambda}](z) = \sum_{k = 0}^{\infty}\omega^{k} \mathcal{C}[k](z)
	\,,
\end{equation}
which can be further expanded in powers of $\tilde{\lambda}^{\dot{\alpha}}$ as
\begin{equation}
	\label{eq:expansion}
	\mathcal{C}[\tilde{\lambda}](z) = \sum_{t_1, t_2 = 0}^{\infty}\omega^{t_1 + t_2}{1 \over t_1! t_2!}(\tilde{\lambda}^{\dot{1}})^{t_1}(\tilde{\lambda}^{\dot{2}})^{t_2} \mathcal{C}[t_1, t_2](z).
\end{equation}
The modes $\mathcal{C}[t_1, t_2](z)$ are eigenstates of Lorentz boosts, which we may take as a convenient basis of chiral algebra generators. Each of our generators will come in towers labelled by multi–indices $t=(t_1,t_2)\in\mathbb Z_{\ge 0}^2$; the labels track collinear weights and are additive in OPE channels. This basis is used in the definition of conformally soft modes \cite{Guevara:2021abz} and, among other features, makes the symmetry algebras appear in guises familiar from the study of 2d conformal field theory \cite{Strominger:2021mtt}. 

\begin{figure}
	\centering
	\begin{tikzpicture}[
		scale=2.2, % twice the previous 1.1
		vertex/.style={circle,fill,inner sep=1.2pt},
		propagate/.style={thick},
		defect/.style={very thick,gray}
		]
		% Left vertical defect (same height as before; scaling handles the 2x)
		\draw[defect] (0,-1.0) -- (0,1.0);
		
		% Horizontal leg length (same base length; scaling makes it 2x visually)
		\def\legX{2.2}
		
		% Five legs: y positions and labels
		\foreach \y/\Lleft/\Lright in {
			0.8 / {$J$} / {$a$},
			0.4 / {$K$} / {$\lambda$},
			0.0 / {$\chi$} / {$\phi$},
			-0.4 / {$\tilde{K}$} / {$\tilde{\lambda}$},
			-0.8 / {$\tilde{J}$} / {$b$}
		}{
			% Vertex with left label
			\node[vertex,label=left:\Lleft] (V) at (0,\y) {};
			% Horizontal leg with right-end label
			\draw[propagate] (V) -- ++(\legX,0) node[right] {\Lright};
		}
	\end{tikzpicture}
	\caption{Bulk-defect coupling}
	\label{fig:defectsource}
\end{figure}
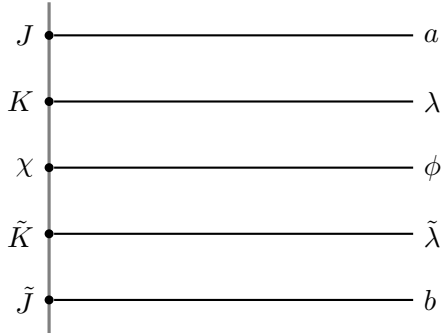

Chiral algebra generators on $\mathbb{CP}^1$ source a corresponding field configuration on twistor space, illustrated schematically in Figure \ref{fig:defectsource}. Their quantum numbers under various symmetries on twistor space are listed in Table \ref{tab:qn}. In addition, all operators $\mathcal{C}[t_1, t_2]$ in a given tower form a highest weight representation of weight $(t_1 + t_2)/2$ of $\mathfrak{sl}(2, \mathbb{C})_+$ and have eigenvalue $(t_1 - t_2)/2$ under its Cartan.  The OPEs are constrained by the requirement that they transform equivariantly under all these symmetries. Notice that the $z$-scaling in the simultaneous dilatation symmetry listed in Table \ref{tab:qn} originates from an element of $\mathfrak{sl}(2, \mathbb{C})_-$ which acts on the projective twistor coordinates as $\lambda = \begin{pmatrix} \lambda_1 \\ \lambda_2 \end{pmatrix} \rightarrow \begin{pmatrix} r^{1/2}\lambda_1 \\ r^{-1/2}\lambda_2 \end{pmatrix}$, such that the full transformation preserves the incidence relation \eqref{eq:incidence} defining spacetime points. In ref.~\cite{FP}, this symmetry was used to restrict the singular terms in the OPEs to be, at worst, double poles. In \S \ref{sec:doublepoles} we show that the OPE of SDSYM has only single poles at any loop order.

\begin{table}[t]
	\centering
	\begin{tabular}{|c|c|c|c|c|c|}
		\hline
		Generator&Field&Scaling Dimension&Spin&Combined Dilatation&Weight  \\
		\hline
		$J[t_1,t_2]$&$a$&$-(t_1+t_2)$&$1-\frac{t_1+t_2}{2}$&$1$&$0$ \\
		\hline
		$K[t_1,t_2]$&$\lambda$&$-(t_1+t_2)-\frac{1}{2}$&$\frac{1}{2}-\frac{t_1+t_2}{2}$&$\frac34$&$\frac14$ \\
		\hline
		$\chi[t_1,t_2]$&$\phi$&$-(t_1+t_2)-1$&$-\frac{t_1+t_2}{2}$&$\frac{1}{2}$&$\frac{1}{2}$ \\
		\hline
		$\tilde{K}[t_1,t_2]$&$\tilde{\lambda}$&$-(t_1+t_2)-\frac32$&$-\frac{1}{2}-\frac{t_1+t_2}{2}$&$\frac14$&$\frac34$ \\
		\hline
		$\tilde{J}[t_1,t_2]$&$b$&$-(t_1+t_2)-2$&$-1-\frac{t_1+t_2}{2}$&$0$&$1$ \\
		\hline
	\end{tabular}
	\caption{Local operators of the 2d chiral algebra, the 6d fields they source, and their quantum numbers. Scaling dimension is the charge under scaling of $\mathbb{R}^4$, spin is the holomorphic 2d conformal weight, combined dilatation is the charge of the operator under simultaneous dilatations $z \rightarrow \frac{z}{r}$ on $\mathbb{CP}^1$ and $x \rightarrow \sqrt{r}x$ on $\mathbb{R}^4$, and weight the charge under an $\hbar$-rescaling. Note that for all operators, $t_1+t_2\geq0$.}
	\label{tab:qn}
\end{table}

\subsection{Tree-level OPEs}
\label{sec:treelevel}
The chiral algebra OPEs corresponding to classical collinear splitting amplitudes may be easily obtained using the method of Koszul duality which is now standard \cite{CPsugra, CP, FPW, Garner:2024tis} (see \cite{PW} for a review). Indeed, one-loop OPEs have also been derived using this method \cite{victor, roland}, and it is straightforward to perform the same computation in SDSYM and verify the results obtained using associativity methods, below.

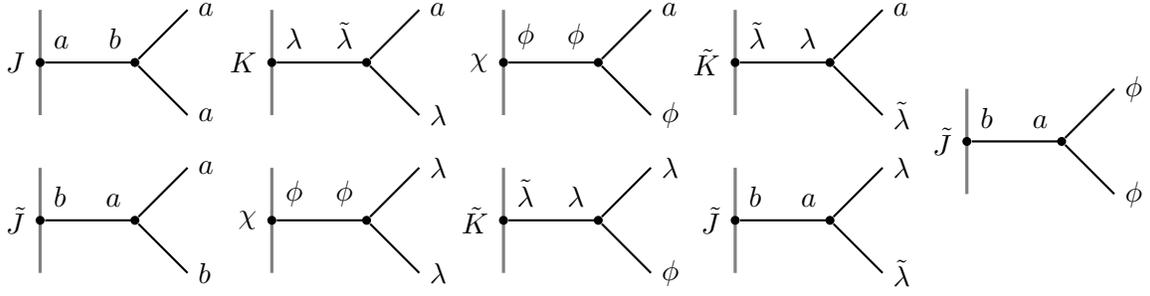
\begin{figure}
	\centering
	\begin{tikzpicture}[
		scale=.7,
		vertex/.style={circle,fill,inner sep=1.2pt},
		propagate/.style={thick}, % switch to: thick,decorate,decoration={snake,amplitude=1.2pt,segment length=4pt}
		defect/.style={very thick,gray}
		]
		
		% Macro:
		% \FDiag{<defect-left label>}{<defect-above-right label>}{<J (above-left) label>}{<A label>}{<B label>}
		\newcommand{\FDiag}[5]{%
			% Shorter defect to allow tight rows
			\draw[defect] (0,-1.0) -- (0,1.0);
			% Defect–propagator vertex with two distinct labels
			\node[vertex,
			label=left:{#1},
			label=above right:{#2}] (Jtilde) at (0,0) {};
			% Horizontal propagator to 3 point vertex
			\draw[propagate] (Jtilde) -- (1.8,0);
			% 3 point vertex with label placed above-left
			\node[vertex,label=above left:{#3}] (J) at (1.8,0) {};
			% Outgoing legs at ±45°, length (1,1)
			\draw[propagate] (J) -- ++(1,1)   node[right] {#4};
			\draw[propagate] (J) -- ++(1,-1)  node[right] {#5};
		}
		
		% Layout spacing
		\def\xstep{4.4}
		\def\ystep{1.5} % tighter vertical spacing as requested
		
		% ---- Top row (4 diagrams) ----
		\begin{scope}[shift={(0*\xstep,\ystep)}]
			\FDiag{$J$}{$a$}{$b$}{$a$}{$a$}
		\end{scope}
		\begin{scope}[shift={(1*\xstep,\ystep)}]
			\FDiag{$K$}{$\lambda$}{$\tilde{\lambda}$}{$a$}{$\lambda$}
		\end{scope}
		\begin{scope}[shift={(2*\xstep,\ystep)}]
			\FDiag{$\chi$}{$\phi$}{$\phi$}{$a$}{$\phi$}
		\end{scope}
		\begin{scope}[shift={(3*\xstep,\ystep)}]
			\FDiag{$\tilde{K}$}{$\tilde{\lambda}$}{$\lambda$}{$a$}{$\tilde{\lambda}$}
		\end{scope}
		
		% ---- Bottom row (4 diagrams) ----
		\begin{scope}[shift={(0*\xstep,-\ystep)}]
			\FDiag{$\tilde{J}$}{$b$}{$a$}{$a$}{$b$}
		\end{scope}
		\begin{scope}[shift={(1*\xstep,-\ystep)}]
			\FDiag{$\chi$}{$\phi$}{$\phi$}{$\lambda$}{$\lambda$}
		\end{scope}
		\begin{scope}[shift={(2*\xstep,-\ystep)}]
			\FDiag{$\tilde{K}$}{$\tilde{\lambda}$}{$\lambda$}{$\lambda$}{$\phi$}
		\end{scope}
		\begin{scope}[shift={(3*\xstep,-\ystep)}]
			\FDiag{$\tilde{J}$}{$b$}{$a$}{$\lambda$}{$\tilde{\lambda}$}
		\end{scope}
		
		% ---- Rightmost, centered between rows ----
		\begin{scope}[shift={(4*\xstep,0)}]
			\FDiag{$\tilde{J}$}{$b$}{$a$}{$\phi$}{$\phi$}
		\end{scope}
		
	\end{tikzpicture}
	\caption{Bulk-defect tree-level Feynman diagrams}
	\label{fig:treeint}
\end{figure}

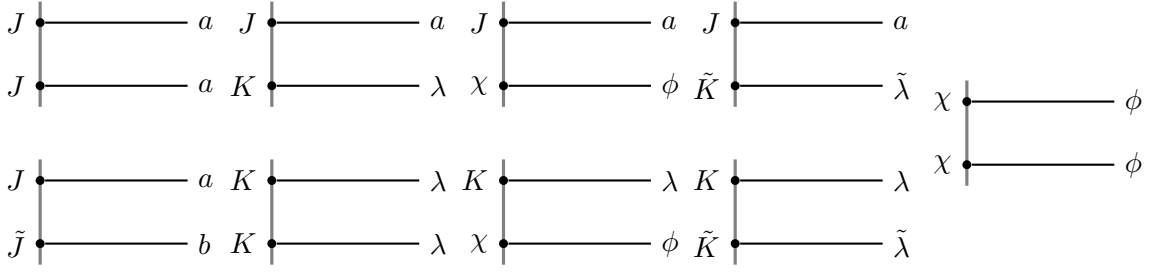
\begin{figure}
	\centering
	\begin{tikzpicture}[
		scale=.7,
		vertex/.style={circle,fill,inner sep=1.2pt},
		propagate/.style={thick}, % for wavy legs: thick,decorate,decoration={snake,amplitude=1.2pt,segment length=4pt}
		defect/.style={very thick,gray}
		]
		
		% Macro:
		% \TwoLegDiag{<upper-vertex left label>}{<upper-end label>}{<lower-vertex left label>}{<lower-end label>}
		\newcommand{\TwoLegDiag}[4]{%
			% Keep same geometry as previous set
			\draw[defect] (0,-1.0) -- (0,1.0);
			\def\yupper{0.6}
			\def\ylower{-0.6}
			\def\legX{2.8}
			
			% Upper vertex (single left label) and leg with right-end label
			\node[vertex,label=left:{#1}] (U) at (0,\yupper) {};
			\draw[propagate] (U) -- ++(\legX,0) node[right] {#2};
			
			% Lower vertex (single left label) and leg with right-end label
			\node[vertex,label=left:{#3}] (L) at (0,\ylower) {};
			\draw[propagate] (L) -- ++(\legX,0) node[right] {#4};
		}
		
		% Spacing (unchanged)
		\def\xstep{4.4}
		\def\ystep{1.5}
		
		% ---- Top row (4 diagrams) ----
		\begin{scope}[shift={(0*\xstep,\ystep)}]
			\TwoLegDiag{$J$}{$a$}{$J$}{$a$}
		\end{scope}
		\begin{scope}[shift={(1*\xstep,\ystep)}]
			\TwoLegDiag{$J$}{$a$}{$K$}{$\lambda$}
		\end{scope}
		\begin{scope}[shift={(2*\xstep,\ystep)}]
			\TwoLegDiag{$J$}{$a$}{$\chi$}{$\phi$}
		\end{scope}
		\begin{scope}[shift={(3*\xstep,\ystep)}]
			\TwoLegDiag{$J$}{$a$}{$\tilde{K}$}{$\tilde{\lambda}$}
		\end{scope}
		
		% ---- Bottom row (4 diagrams) ----
		\begin{scope}[shift={(0*\xstep,-\ystep)}]
			\TwoLegDiag{$J$}{$a$}{$\tilde{J}$}{$b$}
		\end{scope}
		\begin{scope}[shift={(1*\xstep,-\ystep)}]
			\TwoLegDiag{$K$}{$\lambda$}{$K$}{$\lambda$}
		\end{scope}
		\begin{scope}[shift={(2*\xstep,-\ystep)}]
			\TwoLegDiag{$K$}{$\lambda$}{$\chi$}{$\phi$}
		\end{scope}
		\begin{scope}[shift={(3*\xstep,-\ystep)}]
			\TwoLegDiag{$K$}{$\lambda$}{$\tilde{K}$}{$\tilde{\lambda}$}
		\end{scope}
		
		% ---- Rightmost, centered between rows ----
		\begin{scope}[shift={(4*\xstep,0)}]
			\TwoLegDiag{$\chi$}{$\phi$}{$\chi$}{$\phi$}
		\end{scope}
		
	\end{tikzpicture}
	\caption{The gauge variation of these bulk-defect tree diagrams cancel against the gauge variations of the diagrams in~\eqref{fig:treeint} to produce the classical holomorphic collinear splitting functions of 4d $\mathcal{N}=4$ SDSYM.}
	\label{fig:2def}
\end{figure}

One cancels the gauge variation of the source term that couples the 2d chiral algebra generator $\mathcal{C}$ to the corresponding field on twistor space $\mathcal{G}$ (see Figure \ref{fig:2def}),
\begin{equation}
	\sum_{t_1,t_2 \geq 0} \Big( \frac{1}{2 \pi i} \Big) \frac{1}{t_1!\,t_2!}
	\int_{\mathbb{CP}^1} d^2 z \,
	\mathcal{C}_a[t_1,t_2](z)\, \partial_{v^1}^{t_1} \partial_{v^2}^{t_2} \mathcal{G}^a_{\bar z}(z),
	\label{eq:BRSTcoup}
\end{equation} against the gauge variations of the action (for us,~\eqref{eq:supergauge}). The relevant diagrams take simple cubic forms on twistor space, schematically depicted in Figure \ref{fig:treeint}. Gauge invariance of the combined system fixes the singular terms in the OPEs to be: \begingroup\allowdisplaybreaks
\begin{align}
	\label{jjope}
	J_b[r_1,r_2](z)\,J_c[s_1,s_2](w) \sim& \frac{1}{z-w}\; f^{a}_{bc}\; J_a[r_1+s_1, r_2+s_2](w) \\
	\label{kjope}
	K_a^i[r_1, r_2](z) J_b[s_1, s_2](w)\sim & \frac{1}{z-w}f^c_{ab} K^i_c[r_1 + s_1, r_2 + s_2](w) \\
	\label{chijope}
	\chi_a^{ij}[r_1, r_2](z) J_b[s_1, s_2](w)\sim & \frac{1}{z-w} f^c_{ab} \chi^{ij}_c[r_1 + s_1, r_2 + s_2](w) 
	\\
	\label{ktjope}
	\tilde{K}_{ia}[r_1, r_2](z) J_b[s_1, s_2](w)\sim &\frac{1}{z-w}f^c_{ab} \tilde{K}_{ic}[r_1 + s_1, r_2 + s_2](w)
	\\
	\label{jtjope}
	\tilde{J}_a[r_1, r_2](z) J_b[s_1, s_2](w)\sim & \frac{1}{z-w}f^c_{ab} \tilde{J}_{c}[r_1 + s_1, r_2 + s_2](w)  
	\\
	\label{kkope}
	K_a^i[r_1, r_2](z) K_b^j[s_1, s_2](w)\sim & -\frac{1}{z-w}f^c_{ab} \chi^{ij}_c[r_1 + s_1, r_2 + s_2](w)  
	\\
	\label{chikope}
	\chi_a^{ij}[r_1, r_2](z) K_b^k[s_1, s_2](w) \sim & \frac{1}{z-w}\epsilon^{ijkl}f^c_{ab} \tilde{K}_{lc}[r_1 + s_1, r_2 + s_2](w)
	\\
	\label{ktkope}
	\tilde{K}_{ia}[r_1, r_2](z) K_b^j[s_1, s_2](w)\sim &- \frac{1}{z-w}\delta_i^jf^c_{ab} \tilde{J}_{c}[r_1 + s_1, r_2 + s_2](w) 
	\\
	\label{chichiope}
	\chi^{ij}_a[r_1, r_2](z) \chi^{kl}_b[s_1, s_2](w) \sim & \frac{1}{z-w}\epsilon^{ijkl}f^c_{ab} \tilde{J}_{c}[r_1 + s_1, r_2 + s_2](w)
\end{align}\endgroup

It is also convenient to apply Koszul duality directly on $\mathbb{PT}^{3|4}$ to make manifest the $\mathcal{N}=4$ supersymmetry. The chiral algebra is supported on $\mathbb{CP}^{1|4}$, and we package the currents into the single superfield which couples linearly to $\mathcal{A}$:
\begin{equation}
	\label{eq:defect_supercurrent}
	\cJ_a(z,\theta)
	=
	J_a(z)
	+\theta_i K_{a}^i(z)
	+\frac{1}{2} \theta_i\theta_j \chi^{ij}_a(z)
	+\frac{1}{3!}\epsilon^{ijkl}\theta_i\theta_j\theta_k \widetilde K_{al}(z)
	+\frac{1}{4!}\epsilon^{ijkl}\theta_i\theta_j\theta_k\theta_l\,\tilde J_a(z).
\end{equation}
from which we immediately get
\begin{equation}
	\cJ_a[r_1,r_2](z_1,\theta_1)\cJ_b[s_1,s_2](z_2,\theta_2)\sim \frac{1}{z_1-z_2}f_{ab}^c \cJ_c[r_1+s_1, r_2+s_2](z_1,\theta_1+\theta_2),
\end{equation}
whose component expansion recovers eqs.~\eqref{jjope}--\eqref{chichiope}. The action of the Poincar\'e supercharges ($Q_{J,\alpha}, \bar{Q}^J_{\dot\alpha}$) with $\{\alpha,\dot\alpha\}\in\{1,2\}$ and $J\in\{1,2,3,4\}$ can be written:
\begin{align}
	[Q_{J,\alpha},\cJ_a[r_1,r_2](z,\theta)]&=-\theta_Jz^{\alpha-\frac32}\cJ_a[r_1,r_2](z,\theta) \\
	[\bar{Q}^J_{\dot\alpha},\cJ_a[r_1,r_2](z,\theta)]&=\big((2-\dot\alpha)r_1+(\dot\alpha-1) r_2\big)\frac{\partial}{\partial \theta_J}\cJ_a[r_1+\dot\alpha-2,r_2-\dot\alpha+1](z,\theta)
\end{align}
In terms of the component fields, we make use of the action
\begin{align}
\label{eq: Qaction}
	Q_{J,\alpha}J_a[r_1,r_2](z)&=0, \\
	Q_{J,\alpha}K_a^i[r_1,r_2](z)&=\delta_J^iz^{\alpha-\frac32}J_a[r_1,r_2](z), \\
	Q_{J,\alpha}\chi^{ij}_a[r_1,r_2](z)&=\delta^j_J z^{\alpha-\frac32}K^i_a[r_1,r_2](z)-\delta^i_Jz^{\alpha-\frac32}K^j_a[r_1,r_2](z)\\
	Q_{J,\alpha}\tilde K_{ia}[r_1,r_2](z) &=\tfrac{1}{2}\epsilon_{Jijk}z^{\alpha-\frac32}\chi^{jk}_a[r_1,r_2](z)\\
	Q_{J,\alpha}\tilde J_a[r_1,r_2](z)&=\delta^i_Jz^{\alpha-\frac32}\tilde K_{ia}[r_1,r_2](z)
\end{align} 
The supercharges of opposite chirality lower the mode of the primary operator in addition to lowering the $\theta-$degree. It will turn out to be slightly more expedient at higher loop orders to obtain the OPEs of the highest-weight components of the supermultiplet with associativity then repeatedly apply $Q$, rather than working with the supermultiplet OPEs directly.

For previous work applying these supersymmetry transformations to celestial chiral algebra OPEs, see \cite{Tropper}. Note also \cite{Jiang1, Jiang2}, who recovered these tree-level OPEs by studying four dimensional amplitudes in the celestial basis and computing their collinear limits. 

\subsection{All-loop OPEs}

Using the quantum numbers in Table \ref{tab:qn}, we can deduce the form of the OPE of the superfield $\mathcal{J}$ for $m \geq 1$:\begingroup\allowdisplaybreaks
\begin{multline}
	\label{eq:superOPE}
	\cJ_a[t](z_1,\theta_1)\cJ_b[r](z_2,\theta_2)  \sim  \frac{1}{z_1-z_2}  \overset{\sum k_j = p_m}{\sum_{ m \geq 1 \quad k_j^i \geq 0}}   \hbar^m \overset{(m)}{\underset{(t,r)}{f}}[k_1,...,k_{m+1}]^{d_1 \cdot \cdot \cdot d_{m+1}}_{ab}
    \\
    \times:\prod_{j=1}^{m+1} \cJ_{d_j}[k_j]:(z_1,\theta_1+\theta_2)  
\end{multline}
where we have used the definitions
\begin{equation*}
	t \equiv (t^1,t^2) \quad \quad r \equiv (r^1,r^2) \quad \quad s \equiv (s^1,s^2)
\end{equation*} and $p_m := t+r-m$. In the sum, $k_j^i$ denotes the $i = 1,2$-component of the tuple $k_j$.

In order to avoid ambiguity, we specify here that the component expansion of the normally ordered product on the right hand side of eq.~\eqref{eq:superOPE} is such that its $\theta^4$-component is the composite field made entirely of $\tilde J'$s, i.e.~$:\prod_{j=1}^{m+1} \tilde{J}_{d_j}[k_j]:$. All other lower $\theta$-degree components can be obtained by using the chiral supercharge action~\eqref{eq: Qaction} and the Leibniz rule. One can check that performing these operations leads to the component composite fields that have the correct scaling under combined dilatation (Table \ref{tab:qn}). For later use in our form factor computations, we write these all-loop OPEs for the component fields in Appendix \ref{app:components}.

We can also decompose the coefficient $f$ into numerical and Lie-algebraic factors, via: 
\begin{equation}
	\underset{(t,r)}{\overset{(m)}{f}}[k_1,...,k_{m+1}]_{ab}^{d_1 \cdot \cdot \cdot d_{m+1}} = \underset{(t,r)}{\overset{(m)}{f}}(k_1,...,k_{m+1}) K^{d_1 \cdot \cdot \cdot d_{m+1}}_{ab}
\end{equation}
where we have also defined, as in \cite{FP},
\begin{equation}\label{eq:Kdef}
	\quad \quad K^{d_1 \cdot \cdot \cdot d_{m+1}}_{ab} = -f^{d_1}_{a e_1} K^{e_1 e_2} f^{d_2}_{e_2 e_3} \cdot \cdot \cdot f^{d_m}_{e_{2m-2} e_{2m-1}} K^{e_{2m-1} e_{2m}} f^{d_{m+1}}_{e_{2m} b} 
\end{equation}  in terms of the Killing form and structure constants of the gauge Lie algebra. The associativity equation that we use to solve for this coefficient at $m$ loops takes the following form
\begin{align}
\begin{split}
    \oint_{\lvert w \rvert = 2} dw~\bigg( \oint_{\lvert z \rvert = 1} &dz~\chi^{ij}_a[t](0) K^k_b[r](z)\bigg)K^l_c[s](w) 
    \\
    =& \nonumber\oint_{\lvert z \rvert = 2} dz~ \chi^{ij}_a[t](0)\bigg(\oint_{\lvert z-w \rvert = 1} dw~ K^k_b[r](z) K^l_c[s](w) \bigg) 
    \\
    &- \oint_{\lvert z \rvert = 2} dz~  K^k_b[r](z) \bigg(\oint_{\lvert w \rvert = 1} dw~ \chi^{ij}_a[t](0) K^l_c[s](w)\bigg)
\end{split}
\end{align} \endgroup
which can be massaged into a recursion relation matching eq.~(3.46) of ref.~\cite{FP} as we expect. We restate this recursion relation in Appendix \ref{app:recursion} for completeness. The same equation was first solved with homotopical methods in ref.~\cite{Zeng} (see eq.~(10.11)), where its applicability to SDSYM was already noted. In this case, one could have instead used successive application of the chiral and anti-chiral supercharges instead of associativity to obtain this coefficient; that method turns out to be slightly more tedious, but results in the same closed form expression. 

\section{No double poles}
\label{sec:doublepoles}

In the all-loop superfield OPE \eqref{eq:superOPE}, there is no natural way of inserting higher-order poles that respect the twistorial symmetries of Table \ref{tab:qn}. In particular, no composite of $\mathcal{J}$ has the correct scaling under combined dilatation to admit a double (or higher order) pole.

The absence of double poles in supersymmetric theories was anticipated in ref.~\cite{Ball:2023qim} and proved therein at one loop. The twistorial CAB perspective provides a very simple all-loop proof of this conjecture. Indeed, although it is trivial in the case of SDSYM, let us expand on this question from a component field perspective, which can then be easily adapted to theories with less supersymmetry---and whose OPE coefficients fall under the purview of the solution of ref.~\cite{FP} already. 

Therefore, let us momentarily neglect manifest supersymmetry and work with component fields on $\mathbb{PT}$; see Appendix \ref{app:components} for our detailed notation. The symmetries inherited from $\mathbb{PT}$ summarized in Table \ref{tab:qn} in principle allow for a $Q-$closed double pole in the $J-J$ OPE between two positive gluons: 
\begin{align}
\begin{split}
    J_a[t](z)J_b[r](0)  \supset \overset{\sum k_j = p_m}{\sum_{ m \geq 1 \quad k_j^i \geq 0}}  \hbar^m \bigg[& \frac{1}{z^2} \overset{(m)}{\underset{(t,r)}{\alpha}}[k_1,...,k_{m+1}]^{d_1 \cdot \cdot \cdot d_{m+1}}_{ab}:\prod_{j=1}^{m} \tilde{J}_{d_j}[k_j]:
	\\ &+\frac{1}{z} \overset{(m)}{\underset{(t,r)}{\beta}}[k_1,...,k_{m+1}]^{d_1 \cdot \cdot \cdot d_{m+1}}_{ab}\partial_1:\prod_{j=1}^{m} \tilde{J}_{d_j}[k_j]:\bigg]
    \,.
\end{split}
\end{align}
Let us examine the associativity relation:
\begin{align}
\begin{split}
    \oint_{\lvert w \rvert = 2} dw~w\bigg( \oint_{\lvert z \rvert = 1}&dz~J_a[t](0) J_b[r](z)\bigg)J_c[0](w) 
    \\
    =& \oint_{\lvert z \rvert = 2} dz~J_a[t](0)\bigg(\oint_{\lvert z-w \rvert = 1} dw~w J_b[r](z) J_c[0](w) \bigg)   
    \\
    &- \oint_{\lvert z \rvert = 2} dz~J_b[r](z) \bigg(\oint_{\lvert w \rvert = 1} dw~w J_a[t](0) J_c[0](w)\bigg)
    \,.
\end{split}
\end{align}
The second term on the right hand side vanishes: there are no double poles in the $J_a[t](0)J_c[0](w)$ OPE, since the OPE of $J[0]$ with another $J$ is only non-vanishing at tree level (see eq.~\eqref{jjope}). We can write out the first term on the right hand side, after a change of coordinates, as\begingroup\allowdisplaybreaks
\begin{align}
	\oint_{|z| = 2, |z-w| = 1} dz\, J_a[t](0) &( dw\,w\,J_b[r](z)  J_c[0](w))\ \nonumber\\ &= \oint_{|z| = 2, |w| = 1} dz\,J_a[t](0)(dw\,(w+z)\, J_b[r](z)  J_c[0](w+z) ) ,
\end{align} 
which evaluates to
\begin{equation}
	-f_{bc}^e\oint_{z} z\,dz\,J_a[t](0) J_d[r](z)\,  = f_{bc}^e \overset{\sum k_j = p_m}{\sum_{ m \geq 1 \quad k_j^i \geq 0}}  \overset{(m)}{\underset{(t,r)}{\alpha}}[k_1,...,k_{m}]^{d_1 \cdot \cdot \cdot d_{m}}_{ae}:\prod_{j=1}^{m} \tilde{J}_{d_j}[k_j]:(0),
\end{equation} and after doing the $z$ integral, the left hand side becomes 

\begin{align}
\label{LHS}
\begin{split}
    -\oint_{w}&w\,dw\,\overset{\sum k_j = p_m}{\sum_{ m \geq 1 \quad k_j^i \geq 0}}  \bigg[\overset{(m)}{\underset{(t,r)}{a}}[k_1,...,k_{m+1}]^{d_1 \cdot \cdot \cdot d_{m+1}}_{ab}:J_{d_1}[k_1] \prod_{j=2}^{m+1} \tilde{J}_{d_j}[k_j]: (0)
	\\ 
	& -\overset{(m)}{\underset{(t,r)}{b}}[k_1,...,k_{m+1}]^{d_1 \cdot \cdot \cdot d_{m+1}}_{ab}\delta_i^j:K^i_{d_1}[k_1]\tilde K_{jd_2}[k_2] \prod_{j=3}^{m+1} \tilde{J}_{d_j}[k_j]: (0)
	\\  
	& +\frac18\overset{(m)}{\underset{(t,r)}{b}}[k_1,...,k_{m+1}]^{d_1 \cdot \cdot \cdot d_{m+1}}_{ab}\epsilon_{ijkl}:\chi^{ij}_{d_1}[k_1]\chi^{kl}_{d_2}[k_2] \prod_{j=3}^{m+1} \tilde{J}_{d_j}[k_j]:(0)
	\\
	&+\frac12 \overset{(m)}{\underset{(t,r)}{c}}[k_1,...,k_{m+1}]^{d_1 \cdot \cdot \cdot d_{m+1}}_{ab}:\chi^{ij}_{d_1}[k_1]\tilde{K}_{id_2}[k_2]\tilde K_{jd_3}[k_3] \prod_{j=4}^{m+1} \tilde{J}_{d_j}[k_j]:(0)
	\\
	& +\frac1{24}\overset{(m)}{\underset{(t,r)}{d}}[k_1,...,k_{m+1}]^{d_1 \cdot \cdot \cdot d_{m+1}}_{ab}\epsilon^{ijkl}:\tilde  K_{id_1}[k_1]\tilde{K}_{jd_2}[k_2]\tilde K_{kd_3}[k_3]\tilde K_{ld_4}\prod_{j=5}^{m+1} \tilde{J}_{d_j}[k_j]:(0)
	\\
	&+\overset{(m)}{\underset{(t,r)}{\beta}}[k_1,...,k_{m}]^{d_1 \cdot \cdot \cdot d_{m}}_{ab}\partial_1:\prod_{j=1}^{m} \tilde{J}_{d_j}[k_j]:(0)\bigg]J_c[0](w)\, .
\end{split}
\end{align}
Applying the tree-level OPEs to the term proportional to $\beta$ gives
\begin{equation}
	\nonumber
	\overset{\sum k_j = p_m}{\sum_{ m \geq 1 \quad k_j^i \geq 0}}\overset{(m)}{\underset{(t,r)}{\beta}}[k_1,...,k_{m}]^{e_1d_2 \cdot \cdot \cdot d_{m}}_{ab}f_{e_1c}^{d_1}:\prod_{j=1}^{m} \tilde{J}_{d_j}[k_j]:. 
\end{equation}

The other terms in eq.~\eqref{LHS} can also be simplified in the standard way. We express the normally ordered products of operators as integrals between one primary operator and the normally ordered operator with one fewer primary, at the expense of a $z^{-1}$. The OPE between the single primary and $J_c[0]$ is again a tree-level OPE with no double poles; we extract the coefficient of the residue of the single pole in this tree-level OPE (and nothing else) by using a contour deformation. Evaluating the OPE between the remaining normally ordered product and this residue results in the vanishing of the $c$ and $d$ terms. 

Following this same procedure allows us to simplify the $a$ and $b$ terms with the following results:
\begin{multline}
	\label{eq:abc}
	-\overset{\sum k_j = p_m}{\sum_{ m \geq 1 \quad k_j^i \geq 0}}  \overset{k_1}{\sum_{\gamma=0}}f_{e_2c}^{e_3} f_{e_1e_3}^{d_1}\bigg[\sum_{\pi\in \mathrm{Sh}(1,m-1)}
	\overset{(m)}{\underset{(t,r)}{a}}
	\bigl[\gamma,k_{\pi(1)}-\gamma_{\pi(1)},\ldots,k_{\pi(m)}\bigr]^{e_1e_{\pi(1)}d_{\pi(2)}\cdots d_{\pi(m)}}_{ab}
    \\
	- 4\overset{(m)}{\underset{(t,r)}{b}}[\gamma,k_1-\gamma,k_2...,k_{m}]^{e_1e_2d_2\cdot \cdot \cdot d_{m}}_{ab} 
	+3\overset{(m)}{\underset{(t,r)}{b}}[\gamma, k_1-\gamma,...,k_{m}]^{e_1e_2d_2 \cdot \cdot \cdot d_{m}}_{ab}\bigg]:\prod_{j=1}^{m} \tilde{J}_{d_j}[k_j]:(0).
\end{multline} \endgroup
Expressing $a$ and $b$ coefficients in terms of the ``master'' $f$-coefficients using the formulas in Appendix \ref{app:components}, simplifying, and comparing both sides of the associativity equation results in the bracketed coefficient of eq.~\eqref{eq:abc} vanishing identically, leaving behind only:
\begin{equation}
	\overset{(m)}{\underset{(t,r)}{\beta}}[k_1,...,k_{m+1}]^{e_1d_2 \cdot \cdot \cdot d_{m}}_{ab}f_{e_1c}^{d_1}=\overset{(m)}{\underset{(t,r)}{\alpha}}[k_1,...,k_{m+1}]^{d_1 \cdot \cdot \cdot d_{m}}_{ae_1}f_{bc}^{e_1}.
\end{equation}
Satisfying this equality would require
\begin{align}
	\overset{(m)}{\underset{(t,r)}{\beta}}[k_1,...,k_{m+1}]^{e_1d_2 \cdot \cdot \cdot d_{m}}_{ab}\propto \delta_b^{e_1},\qquad \qquad\qquad \overset{(m)}{\underset{(t,r)}{\alpha}}[k_1,...,k_{m+1}]^{d_1 \cdot \cdot \cdot d_{m}}_{ae_1}\propto\delta_{e_1}^{d_1},
\end{align}
for any Lie algebra, which is not possible given the Lie algebraic dependence of the coefficients~\eqref{eq:Kdef}. So, the coefficients of the putative double pole terms must be $\overset{(m)}{{\beta}}=\overset{(m)}{{\alpha}}=0$.

\section{Form factors in SDSYM for SYM amplitudes}
\label{sec:b2formfactors}
With the all-loop holomorphic collinear splitting functions now in hand, we proceed to use the CAB to compute some simple, interesting form factors in SDSYM. First recall the deformation term that one can add to the self-dual theory \eqref{eq:sdN4} to recover the full gauge theory \cite{Mason, Boels:2006ir}:
\begin{equation}
	S_{\mathrm{full}}
	\;=\; S_\mathrm{SDSYM}
	\;+\; {g^2 \over 12} \int_{\mathbb E^{4|8}} d^4x\, d^8\tilde \theta \;\log\det\!\left(\bar\partial_{L_x} + \cA\big|_{L_x}\right),
	\label{eq:MSB-logdet}
\end{equation} 
where $\mathbb E^{4|8}$ is the Euclidean superspace of $\cN=4$ SYM, with $x^i$ the four bosonic and $\tilde \theta^j$ the eight real fermionic coordinates. Here, $\bar\partial_{L_x}$ is the Dolbeault operator on $L_x\simeq\CP^1$ and $\cA|_{L_x}$ is the restriction of the twistor superconnection to that line. Expanding $\log\det=\tr\log$ in powers of $\cA|_{L_x}$ produces the anti-self-dual towers (MHV vertices) absent from the action \eqref{eq:sdN4}, order-by-order in $g^2$. In this expansion, an N$^k$MHV amplitude of SDSYM corresponds to the term with $k+1$ MHV vertices connected by self-dual propagators. In turn, the N$^k$MHV amplitude can be viewed as a form factor in the self-dual gauge theory with $k+1$ insertions of the deformation operator (integrated over their locations in superspace).

We will compute some simple form factors with a single operator insertion. Because we only insert one operator in four dimensions, the dependence on the spacetime coordinate $x$ is $e^{iPx}$, where $P$ is the total momentum of all external states in the form factor. The amplitude is recovered by integrating over all of spacetime, resulting in a momentum-conserving delta function. For simplicity, the operators will be located at $x=0$, and we will enforce momentum conservation by hand to obtain the corresponding amplitude. The form factors will be denoted by
\begin{equation}
	\langle \mathcal{O} | \mathcal{C}_1[\tilde{\lambda}_1](z_1) \ldots \mathcal{C}_n[\tilde{\lambda}_n](z_n) \rangle\,.
\end{equation} 
The chiral algebra generators in the ket are associated to on-shell asymptotic scattering states.

By expanding the momentum eigenstates in conformally soft modes~\eqref{eq:softmodes}, form factors of the 4d theory are seen to be given by a sum of correlation functions of the chiral algebra
\begin{equation}
	\langle \mathcal{O} | \mathcal{C}_1[\tilde{\lambda}_1](z_1) \ldots \mathcal{C}_n[\tilde{\lambda}_n](z_n) \rangle
	=
	\sum_{\substack{k_1,\dotsc,k_n\geq0 \\ \Delta_1+\dots+\Delta_n=-\Delta_\mathcal{O}}}
	\langle \mathcal{O} | \mathcal{C}_1[k_1](z_1) \ldots \mathcal{C}_n[k_n](z_n) \rangle
	\,,
\end{equation}
where $\Delta_\mathcal{O}$ is the scaling dimension of the operator $\mathcal{O}$. The scaling dimensions $\Delta_i$ of the soft modes $\mathcal{C}_i[k_i]$ depend on the value of $k_i$, as in Table~\ref{tab:qn}. Thus, the requirement that $\Delta_1+\dots+\Delta_n=-\Delta_\mathcal{O}$ provides a constraint on the values of the $k_i$ in the sum, which, in turn, only allows for a finite number of conformally soft modes to contribute to a given process. This requirement on the scaling dimensions follows from the chiral algebra being conformal: correlation functions cannot scale under dilations of spacetime.

The values of the $k_i$ place constraints on the analytic forms of the CA correlators. They dictate the asymptotic behavior of the correlators as $z_i\to\infty$ and the number of powers of $\tilde{\lambda}_i$ that the correlator depends on. In particular, a correlator with a $k_i$-mode must contain $k_i$ powers of $\tilde{\lambda}_i$, and it must have a pole of order $-2\times\text{spin}_i$ at $z_i=\infty$, where $\text{spin}_i$ is the spin of $\mathcal{C}_i[k_i]$ as determined by the spin column in Table~\ref{tab:qn}. This latter requirement follows from little group covariance, after fixing the $k_i$ powers of $\tilde{\lambda}_i$. Lorentz symmetry requires that all $\mathfrak{sl}(2, \mathbb{C})_\pm$ indices are contracted, meaning that any appearance of $z_i$ and $\tilde{\lambda}_i$ must come in the form of spinor products $\spa{i}{j}$ and $\spb{i}{j}$ for some $j$, where
\begin{equation}
	\spa{i}{j} := \epsilon_{\alpha\beta}\lambda_i^{\alpha}\lambda_j^{\beta}
	=z_i-z_j
	\hspace{0.5cm}
	\text{and}
	\hspace{0.5cm}
	\spb{i}{j} := -\epsilon_{\dot{\alpha}\dot{\beta}}\tilde{\lambda}_i^{\dot{\alpha}}\tilde{\lambda}_j^{\dot{\beta}}
	\,.
\end{equation} 
A useful selection rule that follows from these considerations is as follows. A correlation function with an insertion of one soft mode with $k_i \geq 2$, and all others with $k_{j \neq i} = 0$, must vanish: the only possible $\mathfrak{sl}(2, \mathbb{C})_+$-invariant is $[ii]=0$. If $k_i=1$ instead, there are no possible Lorentz invariants at all. For brevity of notation, we will also write $\mathcal{C}[0, 0](z) := \mathcal{C}(z)$ for any generator.

With this understanding of how the quantum numbers of Table~\ref{tab:qn} constrain a CA correlation function, the bootstrap proceeds as follows. Given an $n$-point chiral algebra correlator at some loop order, it only has poles at $z_n=z_i$ of the form $\spa{i}{n}$, when viewed as function of $z_n$. The OPEs \eqref{jjloopope}--\eqref{chichiloopope} determine the residues of these poles as either being a correlator of lower-loop order or having a smaller number of external states. In this way, an ansatz can be written down that has the correct pole structure while also satisfying the constraints outlined above. The ansatz can often be verified by arguing that no other terms could be added to it that satisfy both the correct OPE limits and the constraints from the quantum numbers.

In this section, we will study form factors with a single insertion of the deformation operator and its component field constituents, especially $\textrm{tr}(B^2)$. On spacetime, the operator is given by \cite{Koster:2017fvf, Boels:2006ir}:
\begin{align}
	\label{eq:opbsq}
	\mathcal{O}_{B^2}(x)& := {1 \over 12}\int d^8\tilde \theta \log\det\!\left(\bar\partial_{L_x} + \cA\big|_{L_x}\right) \nonumber\\ 
	& = \tr\left\{\tfrac12{B}_{ab} {B}^{ab}
	+\tfrac12\Phi_{ij}\tilde\Lambda^{ia}\tilde\Lambda_{a}^j+\tfrac1{8}\epsilon^{ijkl}\epsilon^{pqrs}[\Phi_{ij},\Phi_{pq}][\Phi_{kl},\Phi_{rs}]\right\}.
\end{align}
Earlier non-supersymmetric form factors with one $\textrm{tr}(B^2)$ insertion have been computed using the CAB in refs.~\cite{CP, CPassoc, CosQCD, DM1}. These form factors will reproduce the tree-level MHV amplitude for $\cN=4$ SYM. We will then compute one- and two-loop form factors involving the scattering of gluons only. The helicities are fixed by the SDSYM Feynman rules: there can be only a single minus-helicity gluon at one loop and none at two loops. It is well known that the one-minus- and all-plus-helicity configurations vanish to all orders in a supersymmetric gauge theory \cite{Grisaru:1976vm,Grisaru:1977px,Parke:1985pn,Kunszt:1985mg}; our computation explicitly confirms these supersymmetry Ward identities (SWI) at specific loop orders, highlighting subtleties of the CAB along the way. 

A useful sanity check throughout our computations in this section and the next is compatibility of our results with Ward identities. Similar to scattering amplitudes, form factors satisfy SWI \cite{Brandhuber:2011tv}. This provides an easy vanishing result for certain helicity configurations, which we also reproduce by explicit computation. In particular, we can show that form factors with several insertions of $\tr(B^m)$ and having external gluons with the all-plus- and one-minus-helicity configurations vanish to any loop order. That is
\begin{align}
	\label{SWI1minus}
	\la \prod_{i=1}^k\tr\left(B^{m_i}\right)(x_i)|\jt_{a_1}[\lamt_1](z_1)J_{a_2}[\lamt_2](z_2)\cdots J_{a_n}[\lamt_n](z_n)
	\ra
	&= 0
	\\
	\label{SWIallplus}
	\la \prod_{i=1}^k\tr\left(B^{m_i}\right)(x_i)|J_{a_1}[\lamt_1](z_1)J_{a_2}[\lamt_2](z_2)\cdots J_{a_n}[\lamt_n](z_n)
	\ra
	&= 0
\end{align}
for all $k \geq 1$ and $m_i>0$. The argument is straightforward, with details provided in Appendix \ref{app:proof}.

Throughout this section and Section \ref{sec:bcubed}, we will use the shorthand notation
\begin{equation}
	\tr_{\mathfrak{g}}(12\cdots n) := \tr_{\mathfrak{g}}\left(t_{a_1}t_{a_2}\cdots t_{a_n}\right)
\end{equation}
for traces of Lie algebra generators taken over the adjoint representation $\mathfrak{g}$ of the Lie algebra. When the subscript $\mathfrak{g}$ is omitted, the trace is understood to be taken over the fundamental representation instead.

\subsection{Tree level}

Since the operator in eq.~\eqref{eq:opbsq} is of dimension 4, the sum of the scaling dimensions of the inserted conformal primaries must be equal to $-4$, as explained earlier; therefore, at tree level, the contributing conformal primaries are the lowest lying energy modes. We will begin with two-point correlators and build up to $n$-point ones. 

\subsubsection{Two points}

The only tree-level two-point form factor is: 
\begin{align}
	\bk{ \cO_{B^2}|\tilde{J}_a(z_1)\tilde{J}_b(z_2)}&=-K_{ab}\bk{12}^2\label{eq:B2J}\,.
\end{align}
Here, only the $\tfrac12\tr B^2$ part of $\cO_{B^2}$ contributes to the form factor. This is fixed up to an overall normalization by little group scaling \cite{CP}, but an easy Feynman diagram computation fixes the normalization (alternatively, by lifting the composite operators to twistor space as in ref.~\cite{BuCasali}). 

\subsubsection{Three points}

The next set of tree-level form factors to compute are:
\begin{gather}
	\bk{\cO_{B^2} | \tilde J_a(z_1)\tilde J_b(z_2) J_c(z_3)},\label{eq:3J}\\
	\bk{\cO_{B^2} |\tilde J_a(z_1)\tilde K_{bi}(z_2)K_c^j(z_3)},\label{eq:J2K}\\
	\bk{\cO_{B^2}| \tilde J_a(z_1)\chi^{ij}_b(z_2)\chi^{kl}_c(z_3)},\label{eq:J2phi}
	\\\bk{\cO_{B^2}|\chi^{ij}_a(z_1)\tilde{K}_{bk}(z_2)\tilde K_{cl}(z_3)}\label{eq:phi2K}.
\end{gather}
We will evaluate the correlators \eqref{eq:3J}, \eqref{eq:J2K}, and \eqref{eq:J2phi} using the CAB, where, again, only the operator $\tfrac12\tr B^2$ contributes. As $z_3\rightarrow\infty$, the correlator \eqref{eq:3J} acquires a second-order zero, so we simply sum the contributions from the $z_3\rightarrow z_1$ and $z_3\rightarrow z_2$ OPEs:
\begin{align}
	\bk{\cO_{B^2}| \tilde J_a(z_1)\tilde J_b(z_2) J_c(z_3)}&=\frac{f_{ac}^d}{\spa{1}{3}}\bk{\tr B^2| \tilde J_d(z_1)\tilde J_b(z_2)} + 
    \frac{f_{bc}^d}{\spa{2}{3}}\bk{\tr B^2| \tilde J_a(z_1)\tilde J_d(z_2)}\nonumber
	\\&=-f_{ac}^d\delta_{bd}\frac{\bk{12}^2}{\bk{13}}-f_{bc}^d\delta_{ad}\frac{\bk{12}^2}{\bk{23}}\nonumber\\&=-f_{abc}\frac{\bk{12}^3}{\bk{23}\bk{13}}.
\end{align}
Similarly, the form factor~\eqref{eq:J2K} has a simple zero as $z_3\rightarrow \infty$. The pole at $z_1$ has a vanishing residue, so we evaluate the contribution from the pole at $z_2$ as: 
\begin{align}
	\bk{\cO_{B^2}|\tilde J_a(z_1)\tilde K_{bi}(z_2)K_c^j(z_3)}&\nonumber=-\delta^j_i\frac{f_{bc}^d}{\spa{2}{3}} \bk{\tr B^2| \tilde J_a(z_1)\tilde J_d(z_2)}\\&=\delta_i^jf_{bc}^d\delta_{ad}\frac{\bk{12}^2}{\bk{23}}
	\nonumber\\&=\delta_i^jf_{abc}\frac{\bk{12}^2}{\bk{23}}.
	\label{eq:tJKK}
\end{align}
The form factor~\eqref{eq:J2phi} is a little trickier, since it behaves like a constant as $z_2\rightarrow\infty$ or $z_3\rightarrow\infty$ and has a second order pole as $z_1\rightarrow\infty$. The only non-zero $\mathfrak{sl}(2,\mathbb{C})_-$ invariant compatible with these constraints is $\bk{12}\bk{13}/\bk{23}$. Thus, the form factor has a pole at $z_3=z_2$, and the OPE between $\chi^{ij}_b(z_2)$ and $\chi^{kl}_c(z_3)$ fixes the overall normalization: 
\begin{align}
	\bk{\cO_{B^2}| \tilde J_a(z_1)\chi^{ij}_b(z_2)\chi^{kl}_c(z_3)}&\nonumber=\epsilon^{ijkl}f_{bc}^d\bk{\tr B^2| \tilde J_a(z_1)\tilde J_d(z_2)}\\& =-\epsilon^{ijkl}f_{bc}^d\delta_{ad}\frac{\bk{12}^2}{\bk{23}}=-\epsilon^{ijkl}f_{bc}^d\delta_{ad}\frac{\bk{12}\bk{13}}{\bk{23}}
	\nonumber\\&=-\epsilon^{ijkl}f_{abc}\frac{\bk{12}\bk{13}}{\bk{23}}
    \,.
\end{align}
In the third equality, we used Lorentz symmetry to match the correct scaling at infinity and the fact that $\langle 1 3 \rangle \rightarrow \langle 1 2 \rangle$ in the $z_3 \rightarrow z_2$ OPE limit.

Only the operator $\tfrac{1}{2}\tr\Phi\tilde\Lambda^2$ contributes to the form factor \eqref{eq:phi2K}. The kinematic dependence of this form factor is fixed by the behavior of $z_2$ and $z_3$ at infinity up to an overall normalization. However, the behavior is singular at infinity, so OPEs cannot fix the normalization. Therefore, we must fix the normalization via a simple Feynman diagram computation. The result is
\begin{align}
	\bk{\cO_{B^2}|\chi^{ij}_a(z_1)\tilde{K}_{bk}(z_2)\tilde K_{cl}(z_3)}&=\delta_k^{[i}\delta_l^{j]}f_{abc}\bk{23}.
\end{align}

\subsubsection{Four points} 

There are eight non-vanishing four-point form factors in the MHV sector: 
\begingroup \allowdisplaybreaks
\begin{gather}
	\bk{\cO_{B^2} |J_a(z_1)\tilde J_b(z_2)J_c(z_3)\tilde J_d(z_4)}
    \label{eq:4J}
    \\
	\bk{\cO_{B^2} |J_a(z_1)\tilde J_b(z_2)K_{c}^i(z_3)\tilde K_{jd}(z_4)}
    \label{eq:2J2K}
    \\
	\bk{\cO_{B^2} |J_a(z_1)\tilde J_b(z_2)\chi^{ij}_c(z_3)\chi^{kl}_d(z_4)}
    \label{eq:2J2phi}
    \\
	\bk{\cO_{B^2}|J_a(z_1)\chi^{ij}_b(z_2)\tilde{K}_{ck}(z_3)\tilde K_{dl}(z_4)}
    \label{eq:Jphi2tK}
    \\
	\bk{\cO_{B^2} | K_{a}^i(z_1)\tilde K_{jb}(z_2)K_{c}^k(z_3)\tilde K_{ld}(z_4)}
    \label{eq:4K}
    \\
	\bk{\cO_{B^2} | K_{a}^i(z_1)\tilde K_{jb}(z_2)\chi^{kl}_c(z_3)\chi^{mn}_d(z_4)}
    \label{eq:2K2phi}
    \\
	\bk{\cO_{B^2} | K_{a}^i(z_1) K_{b}^j(z_2)\chi^{kl}_c(z_3)\tilde J_d(z_4)}
    \label{eq:2KphiJ}
    \\
	\bk{\cO_{B^2} | \chi^{ij}_a(z_1)\chi_b^{kl}(z_2)\chi^{pq}_c(z_3)\chi^{rs}_d(z_4)}
    \label{eq:4phi}.
\end{gather}

For the form factors \eqref{eq:4J}, \eqref{eq:2J2K}, and \eqref{eq:2J2phi}, only $\tfrac12\tr B^2$ has a non-zero contribution. As $z_1\rightarrow\infty$, each of these three form factors acquires second-order zeros. These form factors are then given by the sum of contributions from the $z_1\rightarrow z_2$, $z_1\rightarrow z_3$, and $z_1\rightarrow z_4$ OPEs: 
\begin{align}
	&\bk{\cO_{B^2}|J_a(z_1)\tilde J_b(z_2)J_c(z_3)\tilde J_d(z_4)}\nonumber
	= \frac{f_{ab}^e}{\bk{12}}\bk{\tfrac{1}{2}\tr B^2 |\tilde J_e(z_2)J_c(z_3)\tilde J_d(z_4)} 
    \\
    &\qquad\qquad+\frac{f_{ac}^e}{\bk{13}}\bk{\tfrac{1}{2}\tr B^2 |\tilde J_b(z_2)J_e(z_3)\tilde J_d(z_4)}\nonumber +\frac{f_{ad}^e}{\bk{14}}\bk{\tfrac{1}{2}\tr B^2 |\tilde J_b(z_2)J_c(z_3)\tilde J_e(z_4)}
    \\&
	\qquad\qquad =-\frac{\bk{24}^3}{\bk{23}\bk{34}}\bigg[\frac{\tr([ab][cd])}{\bk{12}}-\frac{\tr([ac][bd])}{\bk{13}}+\frac{\tr([ad][bc])}{\bk{14}}\bigg]
    \nonumber
    \\&
	\qquad\qquad =\bk{24}^4\bigg[\frac{\tr(abcd)+\tr(adcb)}{\bk{12}\bk{23}\bk{34}\bk{41}}+\frac{\tr(abdc)+\tr(acdb)}{\bk{12}\bk{24}\bk{43}\bk{31}}
    \nonumber
    \\
    &\hspace{9cm}+\frac{\tr(acbd)+\tr(adbc)}{\bk{13}\bk{32}\bk{24}\bk{41}}\bigg],
    \\
    \nonumber
    \\
	&\bk{\cO_{B^2} |J_a(z_1)\tilde J_b(z_2)K_{c}^i(z_3)\tilde K_{jd}(z_4)}=\frac{f_{ab}^e}{\bk{12}}\bk{\tfrac{1}{2}\tr B^2 |\tilde J_e(z_2)K_{c}^i(z_3)\tilde K_{jd}(z_4)} \nonumber
    \\
    &\qquad\qquad+\frac{f_{ac}^e}{\bk{13}}\bk{\tfrac{1}{2}\tr B^2 |\tilde J_b(z_2)K_{e}^i(z_3)\tilde K_{jd}(z_4)}\nonumber +\frac{f_{ad}^e}{\bk{14}}\bk{\tfrac{1}{2}\tr B^2 |\tilde J_b(z_2)K_{c}^i(z_3)\tilde K_{je}(z_4)}
    \\&
	\qquad\qquad =-\delta^i_j\frac{\bk{24}^2}{\bk{34}}\bigg[\frac{\tr([ab][cd])}{\bk{12}}-\frac{\tr([ac][bd])}{\bk{13}}+\frac{\tr([ad][bc])}{\bk{14}}\bigg]\nonumber
    \\&
	\qquad\qquad =\delta^i_j\bk{23}\bk{24}^3\bigg[\frac{\tr(abcd)+\tr(adcb)}{\bk{12}\bk{23}\bk{34}\bk{41}}+\frac{\tr(abdc)+\tr(acdb)}{\bk{12}\bk{24}\bk{43}\bk{31}}
    \nonumber
    \\
    &\hspace{9cm}+\frac{\tr(acbd)+\tr(adbc)}{\bk{13}\bk{32}\bk{24}\bk{41}}\bigg],
    \\
    \nonumber
    \\
	&\bk{\cO_{B^2} |J_a(z_1)\tilde J_b(z_2)\chi^{ij}_c(z_3)\chi^{kl}_d(z_4)}=\frac{f_{ab}^e}{\bk{12}}\bk{\tfrac{1}{2}\tr B^2 |\tilde J_e(z_2)\chi_{c}^{ij}(z_3)\chi_{d}^{kl}(z_4)} \nonumber
    \\&\qquad\qquad+\frac{f_{ac}^e}{\bk{13}}\bk{\tfrac{1}{2}\tr B^2 |\tilde J_b(z_2)\chi_{e}^{ij}(z_3)\chi_{d}^{kl}(z_4)}\nonumber +\frac{f_{ad}^e}{\bk{14}}\bk{\tfrac{1}{2}\tr B^2 |\tilde J_b(z_2)\chi_{c}^{ij}(z_3)\chi_{e}^{kl}(z_4)}
    \\&
	\qquad\qquad =-\epsilon^{ijkl}\frac{\bk{23}\bk{24}}{\bk{34}}\bigg[\frac{\tr([ab][cd])}{\bk{12}}-\frac{\tr([ac][bd])}{\bk{13}}+\frac{\tr([ad][bc])}{\bk{14}}\bigg]
    \nonumber
    \\&
	\qquad\qquad =\epsilon^{ijkl}\bk{23}^2\bk{24}^2\bigg[\frac{\tr(abcd)+\tr(adcb)}{\bk{12}\bk{23}\bk{34}\bk{41}}+\frac{\tr(abdc)+\tr(acdb)}{\bk{12}\bk{24}\bk{43}\bk{31}}
    \nonumber
    \\
    &\hspace{9cm}+\frac{\tr(acbd)+\tr(adbc)}{\bk{13}\bk{32}\bk{24}\bk{41}}\bigg]
    \,,
\end{align}
where the square bracket in the color traces is the commutator $[ab]=ab-ba$.  

The non-vanishing parts of the form factor~\eqref{eq:Jphi2tK} come only from the 4d operator $\tfrac{1}{2}\tr\Phi\tilde\Lambda^2$. As $z_1\rightarrow\infty$, the form factor acquires a second-order zero. Just as we did before, we sum the contributions of the $z_1\rightarrow z_2$, $z_1\rightarrow z_3$, and $z_1\rightarrow z_4$ OPEs:
\begin{align}
	&\bk{\cO_{B^2}|J_a(z_1)\chi^{ij}_b(z_2)\tilde{K}_{ck}(z_3)\tilde K_{dl}(z_4)}\nonumber
	= \frac{f_{ab}^e}{\bk{12}}\bk{\tfrac{1}{2}\tr\Phi\tilde\Lambda^2 |\chi^{ij}_e(z_2)\tilde{K}_{ck}(z_3)\tilde K_{dl}(z_4)} 
    \\
    &\qquad\qquad+\frac{f_{ac}^e}{\bk{13}}\bk{\tfrac{1}{2}\tr\Phi\tilde\Lambda^2 |\chi^{ij}_b(z_2)\tilde{K}_{ek}(z_3)\tilde K_{dl}(z_4)}\nonumber +\frac{f_{ad}^e}{\bk{14}}\bk{\tfrac{1}{2}\tr\Phi\tilde\Lambda^2|\chi^{ij}_b(z_2)\tilde{K}_{ck}(z_3)\tilde K_{el}(z_4)}
    \\&
	\qquad\qquad =\delta^{[i}_k\delta^{j]}_l\bk{34}\bigg[\frac{\tr([ab][cd])}{\bk{12}}-\frac{\tr([ac][bd])}{\bk{13}}+\frac{\tr([ad][bc])}{\bk{14}}\bigg]\nonumber
    \\&
	\qquad\qquad =-\delta^{[i}_k\delta^{j]}_l\bk{23}\bk{24}\bk{34}^2\bigg[\frac{\tr(abcd)+\tr(adcb)}{\bk{12}\bk{23}\bk{34}\bk{41}}+\frac{\tr(abdc)+\tr(acdb)}{\bk{12}\bk{24}\bk{43}\bk{31}}
    \nonumber
    \\
    &\hspace{9cm}+\frac{\tr(acbd)+\tr(adbc)}{\bk{13}\bk{32}\bk{24}\bk{41}}\bigg]
    \,.
\end{align}\endgroup

The operators $\tfrac12\tr B^2$ and $\tfrac{1}{2}\tr\Phi\tilde\Lambda^2$ both contribute to the form factors \eqref{eq:4K} and \eqref{eq:2K2phi}. As $z_1\rightarrow\infty$, each of these acquires simple zeros. Summing contributions of the $z_1\rightarrow z_2$, $z_1\rightarrow z_3$, and $z_1\rightarrow z_4$ poles gives 
\begingroup \allowdisplaybreaks
\begin{align}
	&\bk{\cO_{B^2}|K_{a}^i(z_1)\tilde K_{jb}(z_2)K_{c}^k(z_3)\tilde K_{ld}(z_4)}=\delta^i_j\frac{f_{ab}^e}{\bk{12}}\bk{\tfrac{1}{2}\tr B^2 |\tilde J_e(z_2)K_{c}^k(z_3)\tilde K_{ld}(z_4)} 
    \nonumber
    \\
    &\qquad\qquad+\frac{f_{ac}^e}{\bk{13}}\bk{\tfrac{1}{2}\tr\Phi\tilde\Lambda^2 |\tilde K_{jb}(z_2)\chi^{ik}_{e}(z_3)\tilde K_{ld}(z_4)}\nonumber +\delta^i_l\frac{f_{ad}^e}{\bk{14}}\bk{\tfrac{1}{2}\tr B^2 |\tilde K_{jb}(z_2)K_{c}^k(z_3)\tilde J_{e}(z_4)}
    \\
    &
	\qquad\qquad =-\delta^i_j\delta^k_l\frac{\bk{24}^2}{\bk{12}\bk{34}}\tr([ab][cd])+\delta^{[i}_j\delta^{k]}_l\frac{\bk{24}}{\bk{13}}\tr([ac][bd])+\delta^i_l\delta^k_j\frac{\bk{24}^2}{\bk{14}\bk{2   3}}\tr([ad][bc])
    \nonumber
    \\
    &
	\qquad\qquad =\bk{24}^2\Big(\delta^i_j\delta^k_j\bk{14}\bk{23}+\delta^i_l\delta^k_j\bk{12}\bk{34}\Big)
    \nonumber
    \\&
	\qquad\qquad\qquad\qquad\qquad \times \bigg[\frac{\tr(abcd)+\tr(adcb)}{\bk{12}\bk{23}\bk{34}\bk{41}}+\frac{\tr(abdc)+\tr(acdb)}{\bk{12}\bk{24}\bk{43}\bk{31}}
    \nonumber
    \\
    &\hspace{9cm}+\frac{\tr(acbd)+\tr(adbc)}{\bk{13}\bk{32}\bk{24}\bk{41}}\bigg]
    \\
    \nonumber
    \\
	&\bk{\cO_{B^2}|K_{a}^i(z_1)\tilde K_{jb}(z_2)\chi_{c}^{kl}(z_3)\chi_{d}^{mn}(z_4)}=\delta^i_j\frac{f_{ab}^e}{\bk{12}}\bk{\tfrac{1}{2}\tr B^2 |\tilde J_e(z_2)\chi_{c}^{kl}(z_3)\chi_{d}^{mn}(z_4)} 
    \nonumber
    \\
    &\hspace{7cm}-\epsilon^{iklp}\frac{f_{ac}^e}{\bk{13}}\bk{\tfrac{1}{2}\tr\Phi\tilde\Lambda^2 |\tilde K_{jb}(z_2)\tilde{K}_{pe}(z_3)\chi_{d}^{mn}(z_4)}
    \nonumber 
    \\
    &\hspace{7cm}-\epsilon^{imnp}\frac{f_{ad}^e}{\bk{14}}\bk{\tfrac{1}{2}\tr\Phi\tilde\Lambda^2 |\tilde K_{jb}(z_2)\chi_{c}^{kl}(z_3)\tilde K_{pe}(z_4)}
    \nonumber
    \\
    &
	\qquad\qquad =-\delta^i_j\epsilon^{klmn}\frac{\bk{23}\bk{24}}{\bk{12}\bk{34}}\tr([ab][cd]) +\epsilon^{iklp}\delta^{[m}_j\delta^{n]}_p\frac{\bk{23}}{\bk{13}}\tr([ac][bd])
    \nonumber
    \\
    &\hspace{8cm} +\epsilon^{imnp}\delta^{[k}_j\delta^{l]}_p\frac{\bk{24}}{\bk{14}}\tr([ad][bc])
    \nonumber
    \\&
	\qquad\qquad =\bk{23}\bk{24}\bigg[\Big(\delta^i_j\epsilon^{klmn}\bk{14}\bk{23}-(\delta^k_j\epsilon^{lmni}+\delta^l_j\epsilon^{mnik})\bk{12}\bk{34}\Big)
    \nonumber
    \\
    &\hspace{8cm}\times\frac{\tr(abcd)+\tr(adcb)}{\bk{12}\bk{23}\bk{34}\bk{41}}
    \nonumber
    \\
    &\qquad\qquad\qquad +\Big(\delta^i_j\epsilon^{klmn}\bk{13}\bk{24}+(\delta^m_j\epsilon^{nikl}+\delta^n_j\epsilon^{iklm})\bk{12}\bk{34}\Big)\frac{\tr(abdc)+\tr(acdb)}{\bk{12}\bk{24}\bk{43}\bk{31}}
    \nonumber
    \\
    &\qquad\qquad\qquad-\Big((\delta^m_j\epsilon^{nikl}+\delta^n_j\epsilon^{iklm})\bk{14}\bk{23}+(\delta^k_j\epsilon^{lmni}+\delta^l_j\epsilon^{mnik})\bk{13}\bk{24}\Big)
    \nonumber
    \\
    &\hspace{8cm}\times\frac{\tr(acbd)+\tr(adbc)}{\bk{13}\bk{32}\bk{24}\bk{41}}\bigg].
\end{align}
\endgroup
Similarly, the form factor~\eqref{eq:2KphiJ} acquires a simple zero as $z_1\rightarrow\infty$. So, we sum the contributions from the $z_1\rightarrow z_2$ and $z_1\rightarrow z_3$ OPEs: \begingroup \allowdisplaybreaks
\begin{align}
	&\bk{\cO_{B^2} | K_{a}^i(z_1) K_{b}^j(z_2)\chi^{kl}_c(z_3)\tilde J_d(z_4)}
    \nonumber
    \\
    &\qquad\qquad= -\frac{f_{ab}^e}{\bk{12}}\bk{\tfrac12\tr B^2|\chi^{ij}_e(z_2)\chi^{kl}_c(z_3)\tilde J_d(z_4)}-\epsilon^{iklm}\frac{f_{ac}^e}{\bk{13}}\bk{\tfrac12\tr B^2|K^j_b(z_2)\tilde K_{me}(z_3)\tilde J _d(z_4)}
    \nonumber
    \\
    &\qquad \qquad=\epsilon^{ijkl}\frac{\bk{34}}{\bk{23}}\bigg[\frac{\bk{24}}{\bk{12}}\tr([ab][cd])-\frac{\bk{34}}{\bk{13}}\tr([ac][bd])\bigg]
    \nonumber
	\\
    &\qquad\qquad=-\epsilon^{ijkl}\bk{14}\bk{24}\bk{34}^4\bigg[\frac{\tr(abcd)+\tr(adcb)}{\bk{12}\bk{23}\bk{34}\bk{41}}+\frac{\tr(abdc)+\tr(acdb)}{\bk{12}\bk{24}\bk{43}\bk{31}}
    \nonumber
    \\
    &\hspace{8cm}+\frac{\tr(acbd)+\tr(adbc)}{\bk{13}\bk{32}\bk{24}\bk{41}}\bigg]
    \,.
\end{align}

Our final four-point tree-level form factor~\eqref{eq:4phi} is slightly more intricate, since it behaves like a constant as any $z_i \rightarrow \infty$. The non-vanishing contributions come from the $\tfrac12\tr B^2$ and $\tfrac{1}{8}\tr \Phi^4$ terms in $\cO_{B^2}$. These form factors need to be evaluated using Feynman diagrams. The results is
\begin{align}
	&\bk{\cO_{B^2} | \chi^{ij}_a(z_1)\chi_b^{kl}(z_2)\chi^{pq}_c(z_3)\chi^{rs}_d(z_4)}
    =\bk{\tfrac12\tr B^2 | \chi^{ij}_a(z_1)\chi_b^{kl}(z_2)\chi^{pq}_c(z_3)\chi^{rs}_d(z_4)}
    \nonumber
    \\
    &\hspace{8cm}+\bk{\tfrac{1}{8}\tr \Phi^4| \chi^{ij}_a(z_1)\chi_b^{kl}(z_2)\chi^{pq}_c(z_3)\chi^{rs}_d(z_4)}
    \nonumber
	\\
    &
    \hspace{1cm}=-\frac12\bigg[\epsilon^{ijkl}\epsilon^{pqrs}\bigg(\frac{\bk{13}\bk{24}+\bk{14}\bk{23}}{\bk{12}\bk{34}}\bigg)\tr([ab][cd])
    \nonumber
    \\
    &\hspace{6cm} +\epsilon^{ijpq}\epsilon^{klrs}\bigg(\frac{\bk{12}\bk{34}-\bk{14}\bk{23}}{\bk{13}\bk{24}}\bigg)\tr([ac][bd])
    \nonumber
    \\
    &\hspace{6cm}-\epsilon^{ijrs}\epsilon^{klpq}\bigg(\frac{\bk{12}\bk{34}+\bk{13}\bk{24}}{\bk{14}\bk{23}}\bigg)\tr([ad][bc])\bigg]
    \nonumber
    \\
    &\hspace{1.5cm}+\frac{1}{2}\bigg[(\epsilon^{ijpq}\epsilon^{klrs}-\epsilon^{ijrs}\epsilon^{klpq})\tr([ab][cd])
    \nonumber
    \\
    &\hspace{2.5cm}+(\epsilon^{ijkl}\epsilon^{pqrs}-\epsilon^{ijrs}\epsilon^{klpq})\tr([ac][bd])+(\epsilon^{ijkl}\epsilon^{pqrs}-\epsilon^{ijpq}\epsilon^{klrs})\tr([ad][bc])\bigg]
    \nonumber
    \\
    & = \Big(\epsilon^{ijkl}\epsilon^{pqrs}\bk{13}\bk{32}\bk{24}\bk{41}+\epsilon^{ijrs}\epsilon^{klpq}\bk{12}\bk{24}\bk{43}\bk{31}\nonumber+\epsilon^{ijpq}\epsilon^{klrs}\bk{12}\bk{23}\bk{34}\bk{41}\Big)
    \nonumber
    \\
    &\qquad\qquad\qquad \times\bigg[\frac{\tr(abcd)+\tr(adcb)}{\bk{12}\bk{23}\bk{34}\bk{41}}+\frac{\tr(abdc)+\tr(acdb)}{\bk{12}\bk{24}\bk{43}\bk{31}}+\frac{\tr(acbd)+\tr(adbc)}{\bk{13}\bk{32}\bk{24}\bk{41}}\bigg]
    \,.
\end{align}

\subsubsection{Higher-point form factors and MHV amplitudes}
\label{sec:npttrees}
All $n\geq 5$-point form factors not involving the scattering of positive-helicity gluons are
\begin{gather}
    \label{kkchichichi}
	\bk{\cO_{B^2} | K_{a}^i(z_1) K_{b}^j(z_2)\chi^{kl}_c(z_3)\chi^{pq}_d(z_4)\chi^{rs}_e(z_5)}
    \\
	\bk{\cO_{B^2} | K_{a}^i(z_1) K_{b}^j(z_2)K^{k}_c(z_3)\chi^{pq}_d(z_4)\tilde K_{re}(z_5)}
    \\
	\bk{\cO_{B^2} | K_{a}^i(z_1) K_{b}^j(z_2)K^{k}_c(z_3)K^l_d(z_4)\tilde J_e(z_5)}
    \\
	\bk{\cO_{B^2} | K_{a}^i(z_1) K_{b}^j(z_2)K^{k}_c(z_3)K^l_d(z_4)\chi^{pq}_e(z_5)\chi^{rs}_f(z_6)}
    \\
	\bk{\cO_{B^2} | K_{a}^i(z_1) K_{b}^j(z_2)K^{k}_c(z_3)K^l_d(z_4)K^p_e(z_5)\tilde K_{qf}(z_6)}
    \\
	\bk{\cO_{B^2} | K_{a}^i(z_1) K_{b}^j(z_2)K^{k}_c(z_3)K^l_d(z_4)K_e^p(z_5)K_f^q(z_6)\chi^{rs}_g(z_7)}
    \\
    \label{kkkkkkkk}
	\bk{\cO_{B^2} | K_{a}^i(z_1) K_{b}^j(z_2)K^{k}_c(z_3)K^l_d(z_4)K_e^p(z_5)K_f^q(z_6)K_g^r(z_7)K_h^s(z_8)}.
\end{gather}\endgroup
These form factors can be computed using the CAB, with the lower-point correlators necessary for the computations given in the previous subsections. The remaining tree-level form factors of $\cO_{B^2}$ are then obtained by including additional insertions of $J$ into the correlators \eqref{kkchichichi}--\eqref{kkkkkkkk} and the lower-point correlators. All other types of insertions vanish simply because such form factors cannot be drawn with the SDSYM Feynman rules. 

The $n$-point form factors resulting from inserting the additional $J$'s comprise the entire tree-level MHV sector of SDSYM. A simple inductive argument can be given to compute these form factors by considering an OPE between a single $J(z_i)$ and the other insertions inside the correlation functions \cite{CP}. We exemplify this argument for the form factor involving a pair of opposite helicity gauginos, a minus-helicity gluon, and $n-3$ positive-helicity gluons. We claim that the form factor is given by
\begin{equation}
	\bk{\cO_{B^2}|\tilde{J}_{a_1}(z_1)K^i_{a_2}(z_2)\tilde K_{ja_3}(z_3)J_{a_4}(z_4)\cdots J_{a_n}(z_n)}=\delta^i_j\underset{\sigma\in S_n/\mathbb{Z}_n} \sum\frac{\bk{12}\bk{13}^2}{\bk{\sigma_1\sigma_2}\cdots\bk{\sigma_n\sigma_1}}\tr(\sigma_1\cdots \sigma_n)
    \,.
	\label{eq:npt}
\end{equation}
When $n=3$, eq.~\eqref{eq:npt} agrees with the three-point correlator \eqref{eq:tJKK}. Now consider the correlation function on the left-hand side of eq.~\eqref{eq:npt} as a function of $z_n$. The correlator has a zero of order two at $z_n=\infty$, and it only has simple poles at $z_n=z_i$. So, it equals the sum of the singular terms arising from the tree-level OPEs \eqref{jjope}--\eqref{chichiope} as $z_n\to z_i$ for each $i$. If we consider only the color-ordered form factor, defined as the kinematic factor that multiplies the color trace $\tr(12\cdots n)$, then there are only poles at $z_n=z_{n-1}$ and $z_n=z_1$. Thus, the color-ordered form factor is given by
\begin{align}
    &\bk{\cO_{B^2}|\tilde{J}(z_1)K^i(z_2)\tilde K_j(z_3)J(z_4)\cdots J(z_n)}
    \nonumber
    \\
    &\hspace{1cm}=
    \left( \frac{1}{\spa{n-1,}{n}}+\frac{1}{\spa{n}{1}} \right) 
    \bk{\cO_{B^2}|\tilde{J}(z_1)K^i(z_2)\tilde K_j(z_3)J(z_4)\cdots J(z_{n-1})}\nonumber
    \\
    &\hspace{1cm}=\frac{\spa{n-1,}{1}}{\spa{n-1,}{n}\spa{n}{1}}\bk{\cO_{B^2}|\tilde{J}(z_1)K^i(z_2)\tilde K_j(z_3)J(z_4)\cdots J(z_{n-1})}\nonumber
    \\
    &\hspace{1cm}=\frac{\spa{n-1,}{1}}{\spa{n-1,}{n}\spa{n}{1}}
    \frac{\spa{1}{2}\spa{1}{3}^2}{\spdenom{n-1,}}\nonumber
    \\
    &\hspace{1cm}=
    \frac{\spa{1}{2}\spa{1}{3}^2}{\spdenom{n}}
    \,,
\end{align}
where we assumed eq.~\eqref{eq:npt} holds for $n-1$ points in going from the second equality to the third one. Therefore, the result \eqref{eq:npt} holds by induction.
 
The entire MHV sector is obtained by applying this induction argument to the $n$-point form factors built from appending an arbitrary number of positive-helicity gluons to the lower-point ones above. These form factors can then be repackaged into the super form factor with the external states being the super field \eqref{eq:defect_supercurrent}:
\begin{equation}
\la\cO_{B^2}|\mathcal{J}_{a_1}(z_1)\cdots\mathcal{J}_{a_n}(z_n)\ra
	=\sum_{\sig\in S_n/\mathbb Z_n}
	\frac{\delta^{(8)}(\sum_{i=1}^n \lambda_i^{\alpha}\eta_i^{A})}
	{\langle\sigma_1\sigma_2\rangle\langle\sigma_2\sigma_3\rangle\cdots\langle \sigma_n\sigma_1\rangle}\tr(\sig_1\sig_2\cdots\sig_n)
    \,.
\end{equation}
The above super form factor is the well-known SDSYM MHV superamplitude of Nair \cite{Nair:1988} in the limit that the operator momentum is sent to zero.

\subsection{One loop} 
We will show that most one-loop form factors of $\cO_{B^2}$ vanish using the CAB. We only discuss form factors that can be drawn with the Feynman rules of SDSYM \eqref{eq:sdN4} and are compatible with the $\mathfrak{sl}(2,\mathbb{C})_+$ selection rules discussed in the introduction to the section. We begin at two points.

\subsubsection{Two points}
\begin{figure}[t]
	\centering
	\begin{tikzpicture}
		\begin{feynman}
			
			\def\opradius{0.22}
			
			%========================================
			% Diagram 1
			%========================================
			\vertex (o1) at (0,0);
			\vertex [dot] (v1) at (2,0);
			\vertex[right] (a1) at (4,1.2) {$\phi$};
			\vertex[right] (a2) at (4,-1.2) {$\phi$};
			
			% external scalar legs
			\diagram*{
				(v1) -- (a1),
				(v1) -- (a2),
			};
			
			% operator
			\draw (o1) circle (\opradius);
			\draw ($(o1)+(-0.10,-0.10)$) -- ($(o1)+(0.10,0.10)$);
			\draw ($(o1)+(-0.10,0.10)$) -- ($(o1)+(0.10,-0.10)$);
			
			% attachment points on operator circle
			\coordinate (o1up) at ($(o1)+({\opradius*cos(55)},{\opradius*sin(55)})$);
			\coordinate (o1dn) at ($(o1)+({\opradius*cos(-55)},{\opradius*sin(-55)})$);
			
			% gauge propagators as plain lines
			\draw (o1up) to[out=85,in=100,looseness=1.5]
			node[pos=0.1, below right] {$B$}
			node[pos=0.92, left] {$A$} (v1);
			
			\draw (o1dn) to[out=-85,in=-100,looseness=1.5]
			node[pos=0.1, above right] {$B$}
			node[pos=0.92, left] {$A$} (v1);
			
			%========================================
			% Diagram 2
			%========================================
			\vertex (o2) at (7,0);
			\vertex [dot] (vt) at (9,0.9);
			\vertex [dot] (vb) at (9,-0.9);
			\vertex[right] (st) at (11,0.9) {$\phi$};
			\vertex[right] (sb) at (11,-0.9) {$\phi$};
			
			% external scalar legs
			\diagram*{
				(vt) -- (st),
				(vb) -- (sb),
			};
			
			% vertical scalar propagator
			\draw (vt) -- node[pos=0, below right] {$\phi$} node[pos=1, above right] {$\phi$} (vb);
			
			% operator
			\draw (o2) circle (\opradius);
			\draw ($(o2)+(-0.10,-0.10)$) -- ($(o2)+(0.10,0.10)$);
			\draw ($(o2)+(-0.10,0.10)$) -- ($(o2)+(0.10,-0.10)$);
			
			% attachment points on operator circle
			\coordinate (o2up) at ($(o2)+({\opradius*cos(35)},{\opradius*sin(35)})$);
			\coordinate (o2dn) at ($(o2)+({\opradius*cos(-35)},{\opradius*sin(-35)})$);
			
			% gauge propagators as plain lines
			\draw (o2up) to[out=60,in=180,looseness=1.5]
			node[pos=0.08, right] {$B$}
			node[pos=0.92, below] {$A$} (vt);
			
			\draw (o2dn) to[out=-60,in=180,looseness=1.5]
			node[pos=0.08, right] {$B$}
			node[pos=0.92, above] {$A$} (vb);
			
		\end{feynman}
	\end{tikzpicture}
	\caption{The Feynman diagrams at one loop with two external scalars that contribute to the form factor \eqref{eq:12phi}. Their sum vanishes.}
	\label{fig:1lfeyn}
\end{figure}
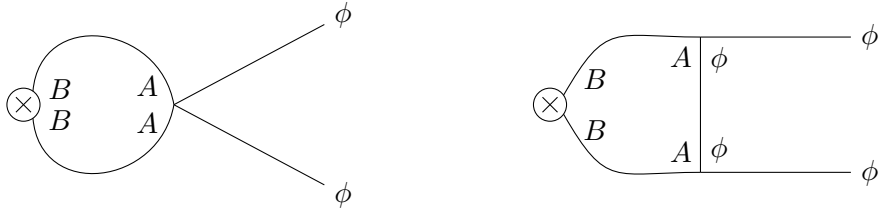

There are three two-point one-loop form factors, all of which vanish. They are
\begin{gather}
	\bk{\cO_{B^2}|\tilde{J}_a[1](z_1)J_b[1](z_2)}\label{eq:12J}
    \\
	\bk{\cO_{B^2}|\tilde{K}_{ai}[1](z_1)K_b^j[1](z_2)}\label{eq:12K}
    \\
    \bk{\cO_{B^2}|\chi^{ij}_a[1](z_1)\chi^{kl}_b[1](z_2)}
    \label{eq:12phi}.
\end{gather}
Only the operator $\tfrac{1}{2}\tr B^2$ contributes non-trivially to the three form factors. The form factor \eqref{eq:12J} acquires a simple zero at $z_2=\infty$, so it is equal to the singular term coming from the one-loop OPE \eqref{jtjloopope} as $z_2\to z_1$:
\begin{align}
	\bk{\cO_{B^2}|\tilde{J}_a[1](z_1)J_b[1](z_2)} &= -\frac{1}{64\pi ^2}\frac{K_{ab}^{cd}}{\bk{12}}\frac{1}{2\pi i}\oint \frac{dw}{z_1-w}\bk{\tfrac{1}{2}\tr B^2|\tilde J_c(z_1)\tilde J_d(w)}
    \nonumber
    \\
	&=-\frac{1}{64\pi^2}\frac{K_{ab}^{cd}}{\bk{12}}\frac1{2\pi i}\oint \frac{dw}{z_1-w}\big(-K_{ab}
	(z_1-w)^2\big)
    \nonumber
    \\
    &=0\,.
\end{align}
For the form factor \eqref{eq:12K}, the only $\mathfrak{sl}(2,\mathbb{C})_-$ invariant consistent with the $z_i\to\infty$ behavior is $\spa{1}{1}=0$. Finally, for the form factor \eqref{eq:12phi}, the combination of $\mathfrak{sl}(2,\mathbb{C})_-$ symmetry and little group scaling implies that it takes the form
\begin{equation}
	\bk{\cO_{B^2}|\chi^{ij}_a[1](z_1)\chi^{kl}_b[1](z_2)}\propto\epsilon^{ijkl}K_{ab}[12]\bk{12}.
\end{equation}
So, the form factor \eqref{eq:12phi} cannot be determined from the OPE between the two $\chi[1]$'s. An explicit Feynman diagram computation shows that the overall constant is, in fact, zero. There are only two non-vanishing diagrams that contribute to the process, seen in Figure \ref{fig:1lfeyn}, and they cancel each other. All other diagrams are identically zero due to color contractions, or because they involve scaleless loop integrals, which vanish in dimensional regularization. 

\subsubsection{Three points}
The following three-point form factors vanish:\begingroup \allowdisplaybreaks
\begin{gather}
	\bk{\cO_{B^2}|\tilde{J}_a[1](z_1)J_b[1](z_2)J_c(z_3)},\label{eq:J1J1J}\\
	\bk{\cO_{B^2}|\tilde{K}_{ia}[1](z_1)K^j_b[1](z_2)J_c(z_3)},\label{eq:K1K1J}\\
	\bk{\cO_{B^2}|\chi^{ij}_a[1](z_1)\chi^{kl}_b[1](z_2)J_c(z_3)},\label{eq:phi1phi1J}\\
	\bk{\cO_{B^2}|\tilde{J}_a(z_1)J_b[1](z_2)J_c[1](z_3)},\label{eq:JJ1J1}\\
	\bk{\cO_{B^2}|K_{a}^i(z_1)\tilde K_{jb}[1](z_2)J_c[1](z_3)},\label{eq:K1KJ1}\\
	\bk{\cO_{B^2}|\tilde{K}_{ia}(z_1)K^j_b[1](z_2)J_c[1](z_3)},\label{eq:KK1J1}\\
	\bk{\cO_{B^2}|\chi^{ij}_a(z_1)\chi^{kl}_b[1](z_2)J_c[1](z_3)},\label{eq:phi1phiJ1}\\
	\bk{\cO_{B^2}|K_{a}^i(z_1)K_{b}^j[1](z_2)\chi^{kl}_c[1](z_3)}.\label{eq:KK1phi1}
\end{gather}\endgroup
It is easy to see that form factors \eqref{eq:J1J1J}--\eqref{eq:phi1phi1J} vanish. As a function of $z_3$, there are only tree-level OPEs between $J(z_3)$ and the other states, and these OPEs give residues that are the two-point one-loop form factors \eqref{eq:12J}--\eqref{eq:12phi}, which we showed are zero. There are no terms regular in these limits, since the form factors \eqref{eq:J1J1J}--\eqref{eq:phi1phi1J} must have second-order zeros at $z_3=\infty$ according to Table \ref{tab:qn}.

There are several ways to see that the form factors \eqref{eq:JJ1J1}--\eqref{eq:phi1phiJ1} also vanish. All of these form factors vanish to first order as $z_3 \rightarrow \infty$, so they are determined by the OPEs between $J[1](z_3)$ and the other states. The tree-level OPEs produce residues proportional to a vanishing two-point one-loop form factor. There remains the possibility of a one-loop OPE as $z_3\rightarrow z_2$. As an example, the one-loop form factor \eqref{eq:KK1J1} in the $z_3\rightarrow z_2$ limit is
\begin{align}
\begin{split}
    \frac{K^{gh}(f_{bg}^df_{ch}^e+f_{cg}^df_{bh}^e)}{\bk{23}}\frac1{2\pi i}\oint\frac{dw}{z_2-w}\Bigg[&\bk{\tfrac12\tr B^2|\tilde{K}_{ia}(z_1)K^j_d(z_2)\tilde J_e(w)}
    \\
    &+\bk{\tfrac12\tr\Phi\tilde\Lambda^2 |\tilde{K}_{ia}(z_1)\chi^{jk}_d(z_2)\tilde{K}_{ie}(w)}\Bigg]
    \,.
\end{split}
\end{align}
The integrand is proportional to $f_{ade}$ which is antisymmetric in $(d,e)$, whereas the combination of structure constants outside the integral is symmetric in $(d,e)$. Thus, the color contraction is zero. The same color contraction occurs for the form factors \eqref{eq:JJ1J1}, \eqref{eq:K1KJ1}, and \eqref{eq:phi1phiJ1}. 

It is easy to see that the form factor \eqref{eq:KK1phi1} is also zero. Indeed, it vanishes to first-order as $z_1\rightarrow\infty$, and the $z_1\rightarrow z_2$ and $z_1\rightarrow z_3$ OPE contributions are proportional to vanishing two-point form factors. 

There is one more possible three-point one-loop form factor allowed by the SDSYM rules: 
\begin{equation}
	\bk{\cO_{B^2}|K_{a}^i[1](z_1)K_{b}^j[1](z_2)\chi^{kl}_c(z_3)}.
    \label{eq:K1K1phi}
\end{equation}
Here, $\mathfrak{sl}(2,\mathbb{C})_+$ scaling implies that the form factor must be divisible by $[12]$. Since the form factor behaves as a constant as each $z_i\rightarrow\infty$, the only $\mathfrak{sl}(2,\mathbb{C})_-$ invariant consistent with this behavior is itself a constant. So,
\begin{equation}
\label{eq:K1K1phi2}
	\bk{\cO_{B^2}|K_{a}^i[1](z_1)K_{b}^j[1](z_2)\chi^{kl}_c(z_3)}\propto\epsilon^{ijkl}f_{abc}[12].
\end{equation}
Since this form factor does not have poles in the $z_i$, a Feynman diagram computation, which we do not perform here, is needed to determine the constant of proportionality. Luckily, we do not need this form factor for the two-loop computations.

\subsubsection{Higher-point form factors and one-loop amplitudes}
\label{sec:1loop4pt}
Let us consider four-point correlators first. Appending an additional $J$ to the form factors \eqref{eq:J1J1J}--\eqref{eq:K1K1phi} gives a vanishing four-point form factor, since it is completely determined by the tree-level OPEs between $J$ and any other state. There are only five form factors to consider:
\begin{gather}
	\bk{\cO_{B^2}|\tilde{K}_{ia}(z_1)K^j_b(z_2)J_c[1](z_3)J_c[1](z_4)},
    \label{eq:12K2J}
    \\
	\bk{\cO_{B^2}|\chi^{ij}_a(z_1)\chi^{kl}_b(z_2)J_c[1](z_3)J_c[1](z_4)},
    \label{eq:12phi2J}
    \\
    \bk{\cO_{B^2}|\chi^{ij}_a(z_1)K^k_b[1](z_2)K^l_c[1](z_3)J_d(z_4)},
    \label{eq:1phiKKJ}
    \\
	\bk{\cO_{B^2}|\chi^{ij}_a(z_1)K^k_b[1](z_2)K^l_c(z_3)J_d[1](z_4)},
    \label{eq:1phiKKJ2}
    \\
	\bk{\cO_{B^2}|K^{i}_a(z_1)K^j_b(z_2)K^k_c[1](z_3)K^l_d[1](z_4)}.
    \label{eq:14K}
\end{gather}

The form factor \eqref{eq:12K2J} is zero, since it vanishes to first order as $z_2\to\infty$ and all OPEs between $K(z_2)$ and any other state produce residues that are vanishing three-point one-loop form factors.
Next, we examine the form factor \eqref{eq:12phi2J} as a function of $z_4$. It cannot have a pole at $z_4=z_1$ or $z_4=z_2$, since the OPEs give residues that are proportional to vanishing three-point one-loop form factors. The contribution from tree-level OPE as $z_4\to z_3$ also vanishes. We are left with taking the one-loop OPE between the two $J[1]$'s \eqref{jjloopope}, giving \begingroup\allowdisplaybreaks
\begin{multline}
    \bk{\cO_{B^2}|\chi^{ij}_a(z_1)\chi^{kl}_b(z_2)J_c[1](z_3)J_c[1](z_4)}
    \\
    \sim -\frac{1}{64\pi^2} 
    \frac{K^{gh}(f_{cg}^df_{dh}^e+f_{dg}^df_{ch}^e)}{\bk{34}} \Big[2\bk{\cO_{B^2}|\chi^{ij}_a(z_1)\chi^{kl}_b(z_2):J_e\tilde J_f:(z_3)}
    \\
    \qquad-2\bk{\cO_{B^2}|\chi^{ij}_a(z_1)\chi^{kl}_b(z_2):K^m_e\tilde K_{mf}:(z_3)} 
    \\
    + \tfrac18\epsilon_{pqrs}\bk{\cO_{B^2}|\chi^{ij}_a(z_1)\chi^{kl}_b(z_2):\chi^{pq}_e\chi^{rs}_{f}:(z_3)}\Big].
\end{multline}\endgroup
These three terms cancel, as verified by inputting the expressions for the tree-level form factors \eqref{eq:2J2phi}, \eqref{eq:2K2phi}, and \eqref{eq:4phi}. Performing a similar OPE analysis for each of the form factors \eqref{eq:1phiKKJ}--\eqref{eq:14K} shows that they depend on the three-point form factor \eqref{eq:K1K1phi2}, which we were able to determine up to a proportionality constant. 

All higher-point form factors only contain additional insertions of $J$. The induction argument given in Section \ref{sec:npttrees} implies that all $n$-point one-loop form factors vanish, except the ones that depend on the four-point form factors \eqref{eq:1phiKKJ}--\eqref{eq:14K}. Hence, we have shown that\begingroup\allowdisplaybreaks
\begin{align}
    \la\cO_{B^2}|\tilde{J}_{a_1}[\lamt_1](z_1),J_{a_2}[\lamt_2](z_2),\cdots,J_{a_n}[\lamt_n](z_n)\ra &= 0
    \\
    \la\cO_{B^2}|\tilde{K}_{i\,a_1}[\lamt_1](z_1),K^j_{a_2}[\lamt_2](z_2),J_{a_3}[\lamt_3](z_3),\cdots,J_{a_n}[\lamt_n](z_n)\ra &= 0
    \\
    \la\cO_{B^2}|\chi^{ij}_{a_1}[\lamt_1](z_1),\chi^{kl}_{a_2}[\lamt_2](z_2),J_{a_3}[\lamt_3](z_3),\cdots,J_{a_n}[\lamt_n](z_n)\ra &= 0
    \,,
\end{align}
which reproduces the one-loop SWI for SDSYM amplitudes
\begin{align}
	\cA^{(1)}_n\left(-,+,...,+\right) &= 0
    \\
	\cA^{(1)}_n\left(-\tfrac{1}{2},+\tfrac{1}{2},+,...,+\right) &= 0
    \\
	\cA^{(1)}_n\left(0,0,+,...,+\right) &= 0\,.
\end{align}\endgroup
In particular, since $\la\cO_{B^2}|-,+,\cdots,+\ra = \la\tfrac{1}{2}\tr(B^2)|-,+,\cdots,+\ra$, we have explicitly confirmed the form factor SWI \eqref{SWI1minus}.

\subsection{Two loops}
We perform a similar analysis at two loops. We only consider two-loop form factors of $\cO_{B^2}$ with all external particles being positive-helicity gluons.

\subsubsection{Two points and renormalization}
The only two-point two-loop correlator that does not vanish due to $\mathfrak{sl}(2,\mathbb{C})_+$ symmetry is
\begin{equation}
	\bk{\cO_{B^2}|J_a[2](z_1)J_b[2](z_2)}=CK_{ab}\spb{1}{2}^2
    \,,
\end{equation} 
for some constant $C$. As in ref.~\cite{CosQCD}, we may add the counterterm $\tfrac{\hbar^2}{2}C\tr(F^2)$ to our operator $\cO_{B^2}$, so that
\begin{equation}
    \label{2loop2J}
	\bk{\cO_{B^2}+\tfrac{\hbar^2}{2}C\tr(F^2)|J_a[2](z_1)J_b[2](z_2)} = 0
    \,.
\end{equation}
The $\hbar^2$ reminds us that it is a two-loop counterterm. Since $\tr(F^2)$ is a total derivative, its contribution vanishes when the operator momentum is taken to zero. The form factor of $\tr(F^2)$ is known for an arbitrary number of gluons and is given by \cite{Dixon:2004za,CosQCD}
\begin{equation}
    \la\tr(F^2)|J[\lamt_1](z_1)\cdots J[\lamt_n](z_n)\ra = 
    \frac{\left(\sum_{1\leq i<j\leq n}\spa{i}{j}\spb{j}{i}\right)^2}{\spdenom{n}}
    \,.
\end{equation}

\subsubsection{Three points}
Three-point form factors include
\begin{align}
    \label{eq:J2J2J}
	\bk{\cO_{B^2}+\tfrac{\hbar^2}{2}C\tr(F^2)|J_a[2](z_1)J_b[2](z_2)J_c(z_3)},
    \\
    \label{eq:J2J1J1}
	\bk{\cO_{B^2}+\tfrac{\hbar^2}{2}C\tr(F^2)|J_a[2](z_1)J_b[1](z_2)J_c[1](z_3)},
\end{align}
and the relabelings $(1\leftrightarrow2)$ and $(1\leftrightarrow3)$ of the latter. The form factor \eqref{eq:J2J2J} vanishes, since it has a second order zero as $z_3\rightarrow\infty$, yet the $z_3\rightarrow z_1$ and $z_3\rightarrow z_2$ OPEs produce residues that are the vanishing two-point two-loop form factor \eqref{2loop2J}. 

The form factor \eqref{eq:J2J1J1} must be proportional to $\spb{1}{2}\spb{1}{3}/\spa{2}{3}$ by $\mathfrak{sl}(2,\mathbb{C})_+$ and $\mathfrak{sl}(2,\mathbb{C})_-$ invariance and the required first-order zero as $z_2\to\infty$ or $z_3\to\infty$. However, the OPE between the two $J[1]$'s determines the residue of the $z_3=z_2$ pole to be zero. The tree-level OPE \eqref{jjope} forces the residue to be the vanishing two-point form factor \eqref{2loop2J}, while the one-loop OPE \eqref{jjloopope} implies that the residue is proportional to $\spb{1}{1}=0$. There is no two-loop OPE between two $J[1]$'s. So, the form factor \eqref{eq:J2J1J1} and its relabelings are zero.

\subsubsection{Higher-point form factors and a two-loop amplitude}
The only new correlation function that appears at four points is
\begin{align}
    \label{2loop4J}
	\bk{\cO_{B^2}+\tfrac{\hbar^2}{2}C\tr(F^2)|J_a[1](z_1)J_b[1](z_2)J_c[1](z_3)J_d[1](z_4)}
    \,.
\end{align}
All others vanish, because appending a $J$ to a lower-point correlator that vanishes results in zero. Viewing the form factor \eqref{2loop4J} as a function of $z_1$, we see that is has a zero at $z_1=\infty$, so it is determined by the OPEs as $z_1\to z_i$. Tree-level OPEs result in a three-point two-loop form factor, all of which vanish. As mentioned before, there is no two-loop OPE between two $J[1]$'s. We are left with taking the one-loop OPE as $z_1\to z_2$, getting\begingroup\allowdisplaybreaks
\begin{multline}
\bk{\cO_{B^2}+\tfrac{\hbar^2}{2}C\tr(F^2)|J_a[1](z_1)J_b[1](z_2)J_c[1](z_3)J_d[1](z_4)}
	\\
    \sim-\frac1{64\pi^2}\frac{2K^{gh}(f_{ag}^ef_{bh}^f+f_{bg}^ef_{ah}^f)}{\bk{12}}\frac1{2\pi i}\oint\frac{dw}{z_1-w}
    \\
    \times\Bigg[\bk{\cO_{B^2}+\tfrac{\hbar^2}{2}C\tr(F^2)|J_{e}(z_1)\tilde{J}_{f}(w)J_c[1](z_3)J_d[1](z_4)}
    \\ 
    +\bk{\cO_{B^2}+\tfrac{\hbar^2}{2}C\tr(F^2)|K^i_{e}(z_1)\tilde{K}_{if}(w)J_c[1](z_3)J_d[1](z_4)}
    \\
    +\tfrac18\epsilon_{ijkl}\bk{\cO_{B^2}+\tfrac{\hbar^2}{2}C\tr(F^2)|\chi^{ij}_{e}(z_1)\chi^{kl}_f(w)J_z[1](z_3)J_d[1](z_4)}\Bigg],
\end{multline}\endgroup
where we have replaced the sum of normally ordered products with a sum of contour integrals. We have already shown in Section \ref{sec:1loop4pt} that the one-loop form factors in the integrand vanish individually. So, the $z_1\to z_2$ limit vanishes. The same is true for the limits $z_1\rightarrow z_3$ and $z_1\rightarrow z_4$. Hence, the form factor \eqref{2loop4J} is zero.

Working inductively as we did in Section \ref{sec:npttrees}, we find that
\begin{equation}
    \label{2loopnJ}
    \bk{\cO_{B^2}+\tfrac{\hbar^2}{2}C\tr(F^2)|J_{a_1}[\lamt_1](z_1)\cdots J_{a_4}[\lamt_n](z_n)}
    = 0
    \,.
\end{equation}
In the limit that the operator momentum is taken to zero, eq.~\eqref{2loopnJ} reduces to the statement that the two-loop all-plus amplitude vanishes,
\begin{gather}
	\cA^{(2)}\big(+,...,+)=0
    \,,
\end{gather}
as is well-known from an application of the SWI \cite{Grisaru:1976vm,Grisaru:1977px,Parke:1985pn,Kunszt:1985mg}.

%%%%%%%%%%%%%%%%%%%%%%%%%%%%%%%%%%
\section{$\tr\left(B^3\right)$ form factors}
\label{sec:bcubed}

In QCD, form factors with a $\tr(F^3)$ insertion contribute to Higgs production via gluon fusion at $\mathcal{O}(1/m_t)$ in an effective theory in which the top-quark mass $m_t$ is taken to be infinite \cite{Wilczek:1977zn,Shifman:1978zn,Inami:1982xt}. The computation of these form factors in $\mathcal{N}=4$ SYM is of particular phenomenological interest, as they exhibit the principle of maximal transcendentality \cite{Brandhuber:2017bkg,Brandhuber:2018xzk,Guo:2022pdw}, which states that the terms of highest transcendental weight match between QCD and $\mathcal{N}=4$ SYM \cite{Kotikov:2001sc,Kotikov:2002ab,Kotikov:2004er,Kotikov:2007cy,Dixon:2017nat}. These form factors have been computed up to two loops for three and four points in $\mathcal{N}=4$ SYM \cite{Brandhuber:2017bkg,Brandhuber:2018xzk,Guo:2022pdw}. 

In this section, we compute form factors with a single insertion of the operator
\begin{equation}
	\tr\left(B^3\right) := \tr\left(B_\mu^{~\nu}B_\nu^{~\rho}B_\rho^{~\mu}\right)
	\,,
\end{equation}
the anti-self-dual part of $\tr(F^3)$, up to three loops. To our knowledge, $\tr(B^3)$ form factors have not been computed at loop-level. Moreover, our results will be the first loop-level $n$-gluon results that contribute to the $\tr(F^3)$ form factors.

The three-point $BAA$ interaction of sdYM enforces that such form factors are only potentially non-vanishing up to three loops, simply because no more loops can be drawn with the Feynman rules. This interaction structure also restricts the helicity configuration of each loop. For the $\tr(B^3)$ form factors, the allowed helicity configurations are three-minus, two-minus, one-minus, and all-plus at tree level, one loop, two loops, and three loops, respectively.

Denote the form factor involving the scattering of $n$-gluons with $l$ of them having negative-helicity as\footnote{The coefficient of $\tfrac{\sqrt{2}}{3!}$ is chosen to give the normalization in eq.~\eqref{F3}.}
\begin{equation}
	\F_{n;\,i_1,i_2,\dotsc,i_l} := 	\la\bthree|\cdots \jt_{a_{i_1}}[\lamt_{i_1}](z_{i_1})\cdots \jt_{a_{i_2}}[\lamt_{i_2}](z_{i_2})\cdots \jt_{a_{i_l}}[\lamt_{i_l}](z_{i_l})\cdots\ra
	\,,
\end{equation}
where we have used ellipses in place of the positive-helicity gluons $J_{a_i}[\lamt_i](z_i)$. As discussed above, 
\begin{equation}
	\F_{n;i_1,i_2,\dotsc,i_l} = 0
\end{equation}
whenever $l\geq4$. The results for $l=0, 1,2,3$ are \footnote{Eqs.~\eqref{F1} and \eqref{F0} are restatements of the SWIs \eqref{SWI1minus} and \eqref{SWIallplus}, respectively.}
\begingroup\allowdisplaybreaks
\begin{align}
	\label{F3}
	\F_{n;\,ijk} &= 
	\hbar^0\sum_{\sig\in S_n/\mathbb{Z}_n}i\frac{\spa{i}{j}^2\spa{j}{k}^2\spa{k}{l}^2}{\spdenomsig{n}}\tr\left(\sig_1\sig_2\cdots\sig_n\right)\ ,
	\\
	\label{F2}
	\F_{n;\,ij}
	&= -\frac{\hbar}{2(4\pi)^2}\sum_{\underset{k,l\neq i,j}{2\leq k<l\leq n}}
	\sum_{\sig\in S_{n;ij;kl}} \frac{\spa{i}{j}^3\spa{j}{k}\spb{k}{l}\spa{l}{i}}{\spdenomsig{n}}\trg{\sig_1\sig_2\cdots\sig_n}\ ,
	\\
	\label{F1}
	\F_{n;\,i} &= 0,
	\\
	\label{F0}
	\F_{n} &= 0.
\end{align}\endgroup
The set of permutations $S_{n;ijkl}\subset S_n/\mathbb{Z}_n$ contains all permutations equivalent up to cycles that preserve the cyclic ordering of $i,j,k,l$ or $j,i,k,l$; this set can be defined as
\begin{equation}
	S_{n;ij;kl} = \bigcup_{\sig\in S(\{1,\dotsc,n\}\setminus\{i,j,k,l\})}\left((i,j,k,l)\shuffle\sig\right)\cup\left((j,i,k,l)\shuffle\sig\right)/\mathbb{Z}_n
	\,,
\end{equation}
where $X\shuffle Y$ denotes the shuffle product between two lists $X$ and $Y$. Also, $S(X)$ is the symmetric group of the set $X$. 

The remainder of this section will outline the computations of eqs.~\eqref{F3}--\eqref{F1}. The computation of $\mathcal{F}_{n;i}$ is an explicit check of the SWI \eqref{SWI1minus}. We cannot verify eq.~\eqref{F0} as a three-loop Feynman diagram computation is needed, which is beyond the scope of this paper. As we have exhaustively illustrated the CAB method in the previous section, we will endeavor to be briefer here.

We will need the following notation
\begin{equation}
	K_{a_1a_2\dots a_n}
	:= K_{a_2b_2}K_{a_3b_3}\cdots K_{a_{n-1}b_{n-1}}
	K_{a_1a_n}^{b_2b_3\dotsc b_{n-1}}
	\,,
\end{equation}
where $K_{a_1a_n}^{b_2b_3\dotsc b_{n-1}}$ is the string of structure constants defined in eq.~\eqref{eq:Kdef}.

%%%%%%%%%%%%%%%%%%%%%%%%%%%%%%%%%%%%%%%%%
\subsection{Tree level}
\label{sec:bcubedtree}

Form factors with a single insertion of $\tr(B^3)$ begin at three points with three minus-heliciy gluons. The form factor can be expressed as a single CA correlator
\begin{equation}
	\label{F33}
	\F_{3;123}=\la\bthree|\jt_{a_1}(z_1)\jt_{a_2}(z_2)\jt_{a_3}(z_3)\ra
	\,.
\end{equation}
Only the softest modes of $\jt_{a_i}[\lamt_i](z_i)$ appear, since $\tr(B^3)$ has dimension 6 and $\jt_{a_i}[0]$ has dimension $-2$. The result is
\begin{equation}
	\F_{3;123}=i\spa{1}{2}\spa{2}{3}\spa{3}{1}
	\,.
\end{equation}
The dependence on angle brackets can be determined by the $\mathfrak{sl}(2,\mathbb{C})_-$ symmetry of the CA, but a simple Feynman diagram computation is needed to compute the overall coefficient. 

The $n$-point form factor $\F_{n;\,ijk}$ is given by a single CA correlator involving only the softest modes
\begin{equation}
	\F_{n;\,ijk} = \la\bthree|\cdots \jt_{a_i}(z_i)\cdots \jt_{a_j}(z_j)\cdots \jt_{a_k}(z_k)\cdots\ra
	\,,
\end{equation}
with ellipses denoting insertions of $J_{a_m}(z_m)$. Assuming that $i,j,k\neq n$ without loss of generality, the $n$-point formula \eqref{F3} then follows by induction on $z_n$ using the argument from Section \ref{sec:npttrees}. The form factor $\F_{3;ijk}$ should be equal to minus the form factor with the insertion of $F_-^3$, since $B=-F_-$, and, indeed, formula \eqref{F3} agrees with the known result given in ref.~\cite{Dixon:2004za}.

Some four-point correlators needed for the computation of the one-loop and two-loop form factors \eqref{F2} and \eqref{F1} are\begingroup\allowdisplaybreaks
\begin{align}
	\label{jtjtkkt}
	&\la\bthree|\jt_{a_1}(z_1)\jt_{a_2}(z_2)K^i_{a_3}(z_3)\tilde{K}_{j\,a_4}(z_4)\ra
	= i\delta_j^iK_{a_1a_2a_3a_4}\frac{\spa{1}{2}\spa{2}{4}\spa{4}{1}}{\spa{3}{4}},
	\\
	\label{jtjtchichi}
	&\la\bthree|\jt_{a_1}(z_1)\jt_{a_2}(z_2)\chi^{ij}_{a_3}(z_3)\chi^{kl}_{a_4}(z_4)\ra = \frac{i}{2}\epsilon^{ijkl}K_{a_1a_2a_3a_4}\frac{\spa{1}{2}}{\spa{3}{4}}
	\left(
	\spa{2}{3}\spa{4}{1} + \spa{2}{4}\spa{3}{1}
	\right)\ ,
	\\
	\label{jtktktchi}
	&\la\bthree|\jt_{a_1}(z_1)\tilde{K}_{i\,a_2}(z_2)\tilde{K}_{j\,a_3}(z_3)\chi^{kl}_{a_4}(z_4)\ra =0
	\,.
\end{align}
The correlator \eqref{jtjtkkt} is computed by taking the tree-level $\tilde{K}K$ OPE \eqref{ktkope}. The other non-vanishing correlator \eqref{jtjtchichi} cannot be computed using the CAB due to scalelessness, as explained in the previous section; a simple Feynman-diagram computation is needed. Finally, the correlation function \eqref{jtktktchi} vanishes because it cannot be drawn with the SDSYM Feynman rules.

We also need the following 5-point correlation functions
\begin{multline}
	\label{jtjtjchichi}
	\la\bthree|\jt_{a_1}(z_1)\jt_{a_2}(z_2)J_{a_3}(z_3)\chi^{ij}_{a_4}(z_4)\chi^{kl}_{a_5}(z_5)\ra
	= i\frac{\epsilon^{ijkl}}{2}\frac{\spa{1}{2}}{\spa{4}{5}}
	\left(
	\spa{1}{4}\spa{2}{5}+\spa{1}{5}\spa{2}{4}
	\right)
	\\
	\times
	\left[
	\frac{\spa{2}{4}}{\spa{2}{3}\spa{3}{4}}K_{a_1a_2a_3a_4a_5}
	+\frac{\spa{4}{5}}{\spa{4}{3}\spa{3}{5}}K_{a_1a_2a_4a_3a_5}
	+\frac{\spa{1}{2}}{\spa{1}{3}\spa{3}{2}}K_{a_1a_3a_2a_4a_5}
	\right],
\end{multline}
\begin{multline}
	\label{jtkktchichi}
	\la\bthree|\jt_{a_1}(z_1)K^m_{a_2}(z_2)\tilde{K}_{n\,a_3}(z_3)\chi^{ij}_{a_4}(z_4)\chi^{kl}_{a_5}(z_5)\ra
	= -\frac{i}{2}
	\delta_n^m\epsilon^{ijkl}K_{a_2a_3a_1a_4a_5}
	\\
	\times
	\frac{\spa{1}{2}}{\spa{2}{3}\spa{4}{5}}
	\left(
	\spa{1}{4}\spa{2}{5}+\spa{1}{5}\spa{2}{4}
	\right),
\end{multline}
\begin{multline}
	\label{jtchichichichi}
	\la\bthree|\jt_{a_1}(z_1)\chi_{i_2j_2}^{a_2}(z_2)\chi^{i_3j_3}_{a_3}(z_3)\chi^{i_4j_4}_{a_4}(z_4)\chi^{i_5j_5}_{a_5}(z_5)\ra
	= -iK_{a_2a_3a_1a_4a_5}\epsilon^{i_2j_2i_3j_3}\epsilon^{i_4j_4i_5j_5}
	\\
	\times
	\left[
	\frac{\spa{1}{4}\spa{5}{2}\spa{3}{1}}{\spa{2}{3}\spa{4}{5}}
	+\frac{1}{2}\frac{\spa{1}{2}\spa{3}{1}}{\spa{2}{3}}
	+\frac{1}{2}\frac{\spa{1}{4}\spa{5}{1}}{\spa{4}{5}}
	\right]
	+ \mathcal{C}(3,4,5)
	\,,
\end{multline}\endgroup
where $+\mathcal{C}(3,4,5)$ instructs one to add the non-trivial cycles of 3,4,5 of the preceding expression. The first one \eqref{jtjtjchichi} is computed by taking the tree-level $\jt J$ and $\chi J$ OPEs and inputting the four-point correlator \eqref{jtjtchichi}. The $\tilde{K}K$ and $\chi K$ OPEs, along with the four-point correlators \eqref{jtjtchichi} and \eqref{jtktktchi}, are used to compute the correlator \eqref{jtkktchichi}. The remaining form factor \eqref{jtchichichichi} requires a Feynman diagram computation.

%%%%%%%%%%%%%%%%%%%%%%%%%%%%%%%%%%%%%%%%%
\subsection{One loop}
\label{sec:bcubed1loop}

\begin{figure}[h]
	\centering
	\begin{equation*}
		\begin{tikzpicture}[baseline={(0,0)},scale=0.75]
			\begin{feynman}
				\vertex[crossed dot] (v1) at (-1,0) {};
				\vertex (v2) at (1,0);
				\vertex (g1) at (-3,0) {\(B_{a_1}(p_1)\)};
				\vertex (g2) at (3,0) {\(B_{a_2}(p_2)\)};
				\diagram*{
					(v1) -- [half right] (v2) -- [half right] (v1),
					(g1) -- (v1),
					(g2) -- (v2),
				};
			\end{feynman}
		\end{tikzpicture}
		\;+\;
		(1\leftrightarrow2)
	\end{equation*}
	\caption{The sum of Feynman diagrams that compute the two-point from factor \eqref{F212}. These diagrams give scaleless integrals, which vanish when dimensionally regulated.}
	\label{fig:1loop2minus}
\end{figure}
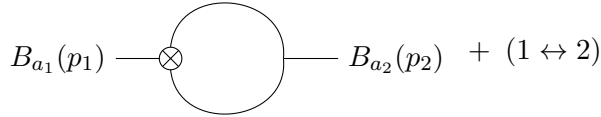

At one loop, the starting point is the two-point two-minus form factor
\begin{equation}
	\label{F212}
	\F_{2;12}=\la\bthree|\jt_{a_1}[1](z_1)\jt_{a_2}[1](z_2)\ra
	\,.
\end{equation}
The correlator on the right hand side is the only one which does not vanish by $\mathfrak{sl}(2,\mathbb{C})_+$ symmetry. In fact, the quantum numbers in Table~\ref{tab:qn} dictate that 
\begin{equation}
	\la\bthree|\jt_{a_1}[1](z_1)\jt_{a_2}[1](z_2)\ra \propto K_{a_1a_2}\spb{1}{2}\spa{1}{2}^3
	\,.
\end{equation}
Because there is no OPE between two $\jt$'s, the CAB cannot determine the constant of proportionality. Luckily, this one-loop form factor is easy to compute with Feynman diagrams, as there are only two of them, shown in Figure~\ref{fig:1loop2minus}. These diagrams vanish, since they are scaleless when dimensionally regulated. So,
\begin{equation}
	\label{F212eqzero}
	\F_{2;12} = \la\bthree|\jt_{a_1}[1](z_1)\jt_{a_2}[1](z_2)\ra = 0
	\,.
\end{equation}

The three-point form factor $\F_{3;12}$ is given by a sum of three CA correlators
\begin{align}
	\begin{split}
		\F_{3;12} = &\la\bthree|\jt_{a_1}[1](z_1)\jt_{a_2}[1](z_2)J_{a_3}(z_3)\ra
		\\
		&+ \la\bthree|\jt_{a_1}[1](z_1)\jt_{a_2}(z_2)J_{a_3}[1](z_3)\ra
		\\
		&+ \la\bthree|\jt_{a_1}(z_1)\jt_{a_2}[1](z_2)J_{a_3}[1](z_3)\ra
		\,.
	\end{split}	
\end{align}
These are the only three that do not vanish by $\mathfrak{sl}(2,\mathbb{C})_+$ symmetry; instead, the OPEs will dictate that they vanish. As a function of $z_3$, the first one vanishes simply due to the residue of the poles at $z_3=z_1$ and $z_3=z_2$ being the vanishing lower point correlator \eqref{F212eqzero}. The other two vanish because the one-loop OPE between $\jt_a[1]J_b[1]$ \eqref{jtjloopope} produces a residue with the normally ordered product $:\jt\jt:$, and
\begin{align}
	\begin{split}
		\la\bthree|\jt_a(z_1):\jt_c\jt_b:(z_2)\ra &=\frac{1}{2\pi i} 
		\int_{\gamma_{z_2}}\frac{dz_3}{z_3-z_2}\la\bthree|\jt_a(z_1)\jt_b(z_2)\jt_c(z_3)\ra
		\\
		&= 
		-\frac{if_{abc}}{2\pi i}\int_{\gamma_{z_2}}dz_3\spa{1}{2}\spa{3}{1}
		\\
		&= 0
		\,.
	\end{split}
\end{align}
The tree-level OPE between $\jt_a[1]J_b[1]$ \eqref{jtjope} produces a two-point correlator that vanishes by $\mathfrak{sl}(2,\mathbb{C})_+$ symmetry. Thus,
\begin{equation}
	\label{F312iszero}
	\F_{3;12} = 0
	\,.
\end{equation}

The first non-vanishing form factor appears at four points, and it is given by a single CA correlation function
\begin{equation}
	\F_{4;12} = \la\bthree|\jt_{a_1}(z_1)\jt_{a_2}(z_2)J_{a_3}[1](z_3)J_{a_4}[1](z_4)\ra
	\,.
\end{equation}
Viewing it as a function of $z_4$, all tree-level OPEs vanish since they give residues that are one of the lower point correlators above. So, the form factor is solely determined by the one-loop OPE between the two $J[1]$'s \eqref{jjloopope}. The result is
\begin{multline}
	\label{jtjtj1j1}
	\la\bthree|\jt_{a_1}(z_1)\jt_{a_2}(z_2)J_{a_3}[1](z_3)J_{a_4}[1](z_4)\ra = -\frac{\hbar}{2(2\pi)^2}\spa{1}{2}^2\frac{\spb{3}{4}}{\spa{3}{4}}
	\\
	\times
	\left(\trg{1234} + \trg{2134}\right)
	\,.
\end{multline}
This form factor gives the seed for a recursion to determine the $n$-point correlator
\begin{multline}
	\label{jtjtj1j1-npt}
	\la\bthree|\jt_{a_1}(z_1)\jt_{a_2}(z_2)J_{a_3}(z_3)\cdots J_{a_k}[1](z_k)\cdots J_{a_l}[1](z_l)\cdots J_{a_n}(z_n)\ra
	\\
	=
	-\frac{\hbar}{2(4\pi)^2}
	\sum_{\sig\in S_{n;12;kl}} \frac{\spa{1}{2}^3\spa{2}{k}\spb{k}{l}\spa{l}{1}}{\spdenomsig{n}}\trg{\sig_1\sig_2\cdots\sig_n}\ ,
\end{multline}
by taking the OPEs of $J_{a_n}(z_n)$ with all other insertions. The full form factor is then given by summing over all possible positions of the two $J[1]$'s:
\begin{equation}
	\F_{n;12} = 
	-\frac{\hbar}{2(4\pi)^2}
	\sum_{3\leq k<l\leq n}
	\sum_{\sig\in S_{n;12;kl}} \frac{\spa{1}{2}^3\spa{2}{k}\spb{k}{l}\spa{l}{1}}{\spdenomsig{n}}\trg{\sig_1\sig_2\cdots\sig_n}
	\,.
\end{equation}
This form factor is the same in any supersymmetric gauge theory, as the terms involving $:K^i\tilde{K}_j:$ and $:\chi_{ij}\chi_{kl}:$ in the OPE between two $J[1]$'s vanish. However, the non-supersymmetric form factor will differ, as it receives a contribution from a double-pole term that supersymmetric theories lack.

We also need the following results for three-point one-loop correlators for the two-loop computation:\begingroup\allowdisplaybreaks
\begin{align}
	\label{jtk1kt1}
	\la\bthree|\jt_{a_1}(z_1)K^i_{a_2}[1](z_2)\tilde{K}_{j\,a_3}[1](z_3)\ra 
	&= 0\ ,
	\\
	\label{jt1k1kt}
	\la\bthree|\jt_{a_1}(z_1)[1]K^i_{a_2}[1](z_2)\tilde{K}_{j\,a_3}(z_3)\ra 
	&= 0\ ,
	\\
	\label{jt1kkt1}
	\la\bthree|\jt_{a_1}(z_1)[1]K^i_{a_2}(z_2)\tilde{K}_{j\,a_3}[1](z_3)\ra 
	&= 0\ ,
	\\
	\label{jtchi1chi1}
	\la\bthree|\jt_{a_1}(z_1)\chi^{ij}_{a_2}[1](z_2)\chi^{kl}_{a_3}[1](z_3)\ra
	&= 0 \ ,
	\\
	\label{jt1chi1chi}
	\la\bthree|\jt_{a_1}[1](z_1)\chi^{ij}_{a_2}[1](z_2)\chi^{kl}_{a_3}(z_3)\ra
	&= 0
	\,.
\end{align}
Eqs.~\eqref{jt1k1kt}, \eqref{jt1kkt1}, and \eqref{jt1chi1chi} can be computed using the OPEs $z_2\to z_3$. For eq.~\eqref{jt1kkt1}, this fact follows from the correlator having a simple zero as $z_2\rightarrow \infty$, according to Table ~\ref{tab:qn}. The other two correlation functions can be determined from the OPEs in the $z_2\to z_3$ limit even though they do not have zeros as $z_2\rightarrow \infty$, since little group scaling requires that
\begin{align}
	\la\bthree|\jt_{a_1}(z_1)[1]K^i_{a_2}[1](z_2)\tilde{K}_{j\,a_3}(z_3)\ra 
	&\propto \delta^i_jf_{a_1a_2a_3}\spb{1}{2}\frac{\spa{1}{2}\spa{1}{3}^2}{\spa{2}{3}},
	\\
	\la\bthree|\jt_{a_1}[1](z_1)\chi^{ij}_{a_2}[1](z_2)\chi^{kl}_{a_3}(z_3)\ra
	&\propto \epsilon^{ijkl}f_{a_1a_2a_3}\spb{1}{2}\frac{\spa{1}{2}^2\spa{1}{3}}{\spa{2}{3}}
	\,.
\end{align}
From these expressions, it is clear that there are no terms regular in the $z_2\to z_3$ limit.

We are resigned to compute eqs.~\eqref{jtk1kt1} and \eqref{jtchi1chi1} by Feynman diagrams, which determine the form factors for the momentum eigenstates 
$$
\la\bthree|\jt_{a_1}[\lamt_1](z_1)K^i_{a_2}[\lamt_2](z_2)\tilde{K}_{j\,a_3}[\lamt_3](z_3)\ra
$$ 
and 
$$
\la\bthree|\jt_{a_1}[\lamt_1](z_1)\chi^{ij}_{a_2}[\lamt_2](z_2)\chi^{kl}_{a_3}[\lamt_3](z_3)\ra\,,
$$ 
(not just the contributions from the conformal modes in eqs.~\eqref{jtk1kt1}--\eqref{jt1chi1chi}). We relate these form factors to those in the conformal basis via the decompositions
\begin{multline}
	\la\bthree|\jt_{a_1}[\lamt_1](z_1)K^i_{a_2}[\lamt_2](z_2)\tilde{K}_{j\,a_3}[\lamt_3](z_3)\ra
	\\
	= \la\bthree|\jt_{a_1}(z_1)K^i_{a_2}[1](z_2)\tilde{K}_{j\,a_3}[1](z_3)\ra 
	\\
	+ \la\bthree|\jt_{a_1}(z_1)[1]K^i_{a_2}[1](z_2)\tilde{K}_{j\,a_3}(z_3)\ra 
	\\
	+ \la\bthree|\jt_{a_1}(z_1)[1]K^i_{a_2}(z_2)\tilde{K}_{j\,a_3}[1](z_3)\ra 
	\\
	= \la\bthree|\jt_{a_1}(z_1)K^i_{a_2}[1](z_2)\tilde{K}_{j\,a_3}[1](z_3)\ra
\end{multline}
and
\begin{multline}
	\la\bthree|\jt_{a_1}[\lamt_1](z_1)\chi^{ij}_{a_2}[\lamt_2](z_2)\chi^{kl}_{a_3}[\lamt_3](z_3)\ra\\
	=\la\bthree|\jt_{a_1}(z_1)\chi^{ij}_{a_2}[1](z_2)\chi^{kl}_{a_3}[1](z_3)\ra
	\\
	+ \la\bthree|\jt_{a_1}[1](z_1)\chi^{ij}_{a_2}[1](z_2)\chi^{kl}_{a_3}(z_3)\ra
	\\
	+ \la\bthree|\jt_{a_1}[1](z_1)\chi^{ij}_{a_2}(z_2)\chi^{kl}_{a_3}[1](z_3)\ra
	\\
	=\la\bthree|\jt_{a_1}(z_1)\chi^{ij}_{a_2}[1](z_2)\chi^{kl}_{a_3}[1](z_3)\ra
	\,.
\end{multline}
The Feynman diagram computation gives zero for both of these form factors.

Some other four-point correlators needed for the two-loop computation are
\begin{align}
	\label{jt1j1kkt}
	\la\bthree|\jt_{a_1}[1](z_1)J_{a_2}[1](z_2)K^i_{a_3}(z_3)\tilde{K}_{j\,a_4}(z_4)\ra
	&= 0\ ,
	\\
	\label{jtj1kkt1}
	\la\bthree|\jt_{a_1}(z_1)J_{a_2}[1](z_2)K^i_{a_3}(z_3)\tilde{K}_{j\,a_4}[1](z_4)\ra
	&=0\ ,
	\\
	\label{jtj1k1kt}
	\begin{split}
		\la\bthree|\jt_{a_1}(z_1)J_{a_2}[1](z_2)K^i_{a_3}[1](z_3)\tilde{K}_{j\,a_4}(z_4)\ra
		&= \delta^i_j\frac{\hbar}{2(4\pi)^2}\frac{\spb{2}{3}}{\spa{2}{3}}\spa{1}{3}\spa{4}{1}
		\\
		&\times\left(\trg{1234}+\trg{1324}\right) \ ,
	\end{split}
	\\
	\label{jt1j1chichi}
	\la\bthree|\jt_{a_1}[1](z_1)J_{a_2}[1](z_2)\chi^{ij}_{a_3}(z_3)\chi^{kl}_{a_4}(z_4)\ra&=0\ ,
	\\
	\label{jtj1chi1chi}
	\begin{split}
		\la\bthree|\jt_{a_1}(z_1)J_{a_2}[1](z_2)\chi^{ij}_{a_3}[1](z_3)\chi^{kl}_{a_4}(z_4)\ra &= -\epsilon^{ijkl}\frac{\hbar}{4(4\pi)^2}\frac{\spb{2}{3}}{\spa{2}{3}}\spa{1}{3}^2
		\\
		&\times\left(\trg{1234}+\trg{1423}\right)
		\,.
	\end{split}
\end{align}
The first two correlators \eqref{jt1j1kkt} and \eqref{jtj1kkt1} vanish, since the OPEs involving $K^i[0]$ are all at tree level, producing residues that are vanishing three-point one-loop correlators. The remaining correlators \eqref{jtj1k1kt}--\eqref{jtj1chi1chi} are computed using OPEs involving $J[1]$. The same tree-level OPE argument above kills off the possible three-point one-loop contributions, while the remaining one-loop $K[1]J[1]$, $\jt[1]J[1]$, and $\chi[1]J[1]$ OPEs will produce residues proportional to the four-point tree-level correlators \eqref{jtjtkkt}--\eqref{jtktktchi}.

Finally, the five-point one-loop correlators needed for the two-loop computation are
\begin{multline}
	\label{jtj1j1kkt}
	\la\bthree|\jt_{a_1}(z_1)J_{a_2}[1](z_2)J_{a_3}[1](z_3)K^i_{a_4}(z_4)\tilde{K}_{j\,a_5}(z_5)\ra
	= -\frac{\hbar}{2(4\pi)^2}\delta^i_j\frac{\spb{2}{3}}{\spa{2}{3}}
	\\
	\times
	\left[
	\left(
	\trg{12345} - \trg{12354} + \trg{13245} - \trg{13254}
	\right)
	\frac{\spa{1}{5}^2}{\spa{4}{5}}
	\right.
	\\
	\left(
	\trg{12354} - \trg{12534} + \trg{13524} - \trg{15324}
	\right)
	\frac{\spa{1}{3}\spa{1}{5}}{\spa{3}{4}}
	\\
	\left.
	\left(
	\trg{13254} - \trg{13524} + \trg{12534} - \trg{15234}
	\right)
	\frac{\spa{1}{2}\spa{1}{5}}{\spa{2}{4}}
	\right],
\end{multline}
\begin{multline}
	\label{jtj1j1chichi}
	\la\bthree|\jt_{a_1}(z_1)J_{a_2}[1](z_2)J_{a_3}[1](z_3)\chi^{ij}_{a_4}(z_4)\chi^{kl}_{a_5}(z_5)\ra 
	= \frac{\hbar}{4(4\pi)^2}\epsilon^{ijkl}\frac{\spb{2}{3}}{\spa{2}{3}}
	\\
	\times
	\left[
	\left(\trg{15234}+\trg{15324}\right)\frac{\spa{1}{2}^2\spa{4}{5}}{\spa{4}{2}\spa{2}{5}}
	\right.
	\\
	+\left(\trg{15432} + \trg{15423}\right)
	\frac{\spa{1}{4}}{\spa{2}{4}\spa{4}{5}}
	\left(\spa{1}{4}\spa{2}{5}+\spa{1}{5}\spa{2}{4}\right)
	\\
	+\left(\trg{12354} + \trg{13254}\right)
	\frac{\spa{1}{5}}{\spa{2}{5}\spa{4}{5}}
	\left(\spa{1}{4}\spa{2}{5}+\spa{1}{5}\spa{2}{4}\right)
	\\
	+\left(
	\trg{15234} - \trg{15243} + \trg{15342} - \trg{15432}
	\right)
	\frac{\spa{1}{4}^2\spa{2}{3}}{\spa{2}{4}\spa{4}{3}}
	\\
	\left.
	+\left(
	\trg{14235} - \trg{14253} + \trg{14352} - \trg{14532}
	\right)
	\frac{\spa{1}{5}^2\spa{2}{3}}{\spa{2}{5}\spa{5}{3}}
	\right]
	\,.
\end{multline}\endgroup
The correlator in eq.~\eqref{jtj1j1kkt} is computed from the OPEs involving the insertion $K^i$. All such OPEs are tree-level and give residues that are the four-point one-loop correlators \eqref{jtjtj1j1-npt} or \eqref{jtj1k1kt}. The correlation function in eq.~\eqref{jtj1j1chichi} is computed from the OPEs involving one of the $J[1]$'s. The $\jt J[1]$ OPE results in the vanishing residue \eqref{jt1j1chichi}, and the $\chi J[1]$ OPEs produce the correlator \eqref{jtj1chi1chi} as their residues. Finally, the one-loop $J[1]J[1]$ OPE \eqref{jjloopope} will generate a residue with terms proportional to the tree-level correlators \eqref{jtjtjchichi}--\eqref{jtchichichichi}. After straightforward algebraic manipulations, the results follow.

%%%%%%%%%%%%%%%%%%%%%%%%%%%%%%%%%%%%%%%%%
\subsection{Two loops}
\label{sec:bcubed2loops}

We conclude our illustrative computations at two loops. As always, we start with the lowest-point contributions and build up inductively to $n$ points. The two-point two-loop form factor
\begin{equation}
	\F_{2;1} = \la\bthree|\jt_{a_1}[2](z_1)J_{a_2}[2](z_2)\ra = 0
\end{equation}
as it cannot have both the correct $z_i\to\infty$ behavior and $\mathfrak{sl}(2,\mathbb{C})_-$ symmetry.

The CA correlation functions contributing to the three-point form factor that do not vanish trivially by a tree-level OPE computation are\begingroup\allowdisplaybreaks
\begin{align}
	\label{jt1j2j1}
	\la\bthree|\jt_{a_1}[1](z_1)J_{a_2}[2](z_2)J_{a_3}[1](z_3)\ra &= 0\ ,
	\\ 
	\label{jt1j1j2}
	\la\bthree|\jt_{a_1}[1](z_1)J_{a_2}[1](z_2)J_{a_3}[2](z_3)\ra &= 0\ ,
	\\
	\label{jt2j1j1}
	\la\bthree|\jt_{a_1}[2](z_1)J_{a_2}[1](z_2)J_{a_3}[1](z_3)\ra &= 0\ ,
	\\
	\label{jtj2j2}
	\la\bthree|\jt_{a_1}(z_1)J_{a_2}[2](z_2)J_{a_3}[2](z_3)\ra &= 0
	\,,
\end{align} though they are all nonetheless zero for the following reasons.
Viewing the first as a function of $z_3$, any tree-level OPE vanishes since the two-point two-loop form factors are zero. We also see that the $\jt[1]J[1]$ one-loop OPE \eqref{jtjloopope} produces correlators that vanish by $\mathfrak{sl}(2,\mathbb{C})_+$ symmetry and produces the factor $\spb{2}{2}=0$. The one-loop $J[2]J[1]$ OPE 
\begin{multline}
	\label{j1j2OPE}
	J_{a_1}[1](z_1)J_{a_2}[2](z_2)\sim-\frac{\hbar}{6(4\pi)^2}\frac{\spb{1}{2}}{\spa{1}{2}}
	\left[
	\left(K_{a_1a_2}^{cd}+2K_{a_1a_2}^{dc}\right):J_c[1](\lamt_2)\jt_d:(z_2)
	\right.
	\\
	+\left(2K_{a_1a_2}^{cd}+K_{a_1a_2}^{dc}\right):J_c(\lamt_2)\jt_d[1]:(z_2)
	+ \left(J_c\jt_d\to-\delta^j_iK^i_c\tilde{K}_{j\,d}\right)
	\\
	\left.
	+ \left(J_c\jt_d\to\tfrac{1}{8}\epsilon_{ijkl}\chi^{ij}_c\chi^{kl}_d\right)
	\right]
\end{multline}\endgroup
gives the vanishing one-loop correlators \eqref{jtk1kt1}--\eqref{jt1chi1chi}. The correlator \eqref{jt1j1j2} vanishes since it is a relabelling of the correlator \eqref{jt1j2j1}.

By the quantum numbers in Table~\ref{tab:qn}, the correlation function \eqref{jt2j1j1} is seen to be proportional to
\begin{equation}
	\la\bthree|\jt_{a_1}[2](z_1)J_{a_2}[1](z_2)J_{a_3}[1](z_3)\ra
	\propto f_{a_1a_2a_3}\spb{1}{2}\spb{1}{3}\frac{\spa{1}{2}^2\spa{1}{3}^2}{\spa{2}{3}^3}
	\,.
\end{equation}
The proportionality constant must be zero, as there cannot be poles of order 2 or greater in a supersymmetric theory. One can check that this is consistent with taking the one-loop $J[1]J[1]$ OPE \eqref{jjloopope}, as it produces a correlation function with $\spb{1}{1}=0$, by $\mathfrak{sl}(2,\mathbb{C})_+$ symmetry.

Finally, the CA correlator \eqref{jtj2j2} is proportional to
\begin{equation}
	\la\bthree|\jt_{a_1}(z_1)J_{a_2}[2](z_2)J_{a_3}[2](z_3)\ra
	\propto f_{a_1a_2a_3}\spb{2}{3}^2\frac{\spa{1}{2}\spa{1}{3}}{\spa{2}{3}}
	\,.
\end{equation}
The appearance of the $z_2=z_3$ pole implies that the proportionality constant can be determined by the $J[2]J[2]$ OPE. The tree-level $J[2]J[2]$ OPE \eqref{jjope} gives a correlator that vanishes due to $\mathfrak{sl}(2,\mathbb{C})_+$ symmetry. The two-loop $J[2]J[2]$ OPE is
\begingroup \allowdisplaybreaks
\begin{multline}
	J_{a_2}[2](z_2)J_{a_3}[2](z_3) \sim -\frac{\hbar^2}{24(4\pi)^4} \frac{\spb{2}{3}^2}{\spa{2}{3}}
	\left[
	\left(-K_{a_2a_3}^{bcd}-K_{a_2a_3}^{cbd}-K_{a_2a_3}^{cdb}\right)
	:J_b\jt_c\jt_d:(z_3)
	\right.
	\\
	\left.
	+ \sum_{\sig\in S(b,c,d)}F_{a_2\sig a_3}
	\left(
	-\delta^j_i :K^i_b\tilde{K}_{j\,c}\jt_d:(z_3)
	+\tfrac{1}{8}\epsilon_{ijkl}:\chi^{ij}_b\chi^{kl}_c\jt_d:(z_3)
	+\tfrac{1}{2}\epsilon_{ijkl}:\chi^{ij}_b\tilde{K}^k_c\tilde{K}^l_d:(z_3)
	\right)
	\right]
	\,.
\end{multline}
\endgroup
One can check that the resulting correlators with these normally ordered products all vanish. The last term vanishes because the tree-level form factor
$$
\la\bthree|\jt_{a_1}(z_1)\chi^{ij}_{a_2}(z_2)\tilde{K}_{k\,a_3}(z_3)\tilde{K}_{l\,a_4}(z_4)\ra
$$
cannot be drawn with Feynman diagrams.

The one-loop $J[2]J[2]$ OPE can be determined from eq.~\eqref{jjloopope}, and it contains many terms of the form $\spb{2}{3}(\lamt_2^{\dot{1}})^{r_1}(\lamt_2^{\dot{2}})^{r_2}(\lamt_3^{\dot{1}})^{s_1}(\lamt_3^{\dot{2}})^{s_2}:O_{1\,c}[t_1,t_2]O_{2\,d}[u_1,u_2]:(z_3)$ with $t_1+t_2+u_1+u_2=2$ and $r_1+r_2+s_1+s_2=2$ such that $r_1\neq r_2$ and $s_1\neq s_2$. The pair $(O_1,O_2)$ can be any of  $(J,\jt), (K^i,\tilde{K}_j), (\chi_{ij},\chi_{kl})$. We have shown already that the one-loop correlators
\begin{equation}
	\la\bthree|\jt_{a_1}(z_1)O_{1\,a_2}[1](z_2)O_{2\,a_3}[1](z_3)\ra
	= 0
\end{equation}
(see eq.~\eqref{F312iszero} and eqs.~\eqref{jtk1kt1}--\eqref{jt1chi1chi}), but
\begingroup \allowdisplaybreaks
\begin{align}
	\label{lamtcorrexpand}
	&\la\bthree|\jt_{a_1}(z_1)O_{1\,a_2}[1](z_2)O_{2\,a_3}[1](z_3)\ra
	=\sum_{t_1+t_2=2}\sum_{u_1+u_2=2}
	\frac{(\lamt_2^{\dot{1}})^{t_1}(\lamt_2^{\dot{2}})^{t_2}
		(\lamt_3^{\dot{1}})^{u_1}(\lamt_3^{\dot{2}})^{u_2}}{t_1!t_2!u_1!u_2!}
    \nonumber
	\\
	&\hspace{4cm}\times
	\la\bthree|\jt_{a_1}(z_1)O_{1\,a_2}[t_1,t_2](z_2)O_{2\,a_3}[u_1,u_2](z_3)\ra
	\,.
\end{align}
\endgroup
Since $\lamt_2$ and $\lamt_3$ are linearly independent complex 2-vectors for generic momenta, 
\begin{equation} \nonumber
(\lamt_2^{\dot{1}})^{t_1}(\lamt_2^{\dot{2}})^{t_2}
(\lamt_3^{\dot{1}})^{u_1}(\lamt_3^{\dot{2}})^{u_2}/t_1!t_2!u_1!u_2! \end{equation} are linearly independent as polynomials. So, each individual correlator on the right hand side of eq.~\eqref{lamtcorrexpand} must be zero. We conclude that eq.~\eqref{jtj2j2} holds and
\begin{equation}
	\F_{3;1}=0\,.
\end{equation}

All four-point CA correlators that contribute to the four-point form factor $\F_{4;1}$ vanish. The nontrivial possibilities are those that do not arise from appending a $J[0]$ to our vanishing three-point correlators. They are
\begin{equation}
	\label{jtj2j1j1}
	\la\bthree|\jt_{a_1}(z_1)J_{a_2}[2](z_2)J_{a_3}[1](z_3)J_{a_4}[1](z_4)\ra
	= 0
	\,,
\end{equation}
plus the relabelings $2\leftrightarrow3$ and $2\leftrightarrow4$, and
\begin{equation}
	\label{jt1j1j1j1}
	\la\bthree|\jt_{a_1}[1](z_1)J_{a_2}[1](z_2)J_{a_3}[1](z_3)J_{a_4}[1](z_4)\ra
	= 0
\end{equation}
View the correlator in eq.~\eqref{jtj2j1j1} as a function of $z_4$. Clearly, any tree-level OPE will result in a lower point two-loop correlation function that is zero. That leaves the one-loop $J[1]J[1]$ and $J[2]J[1]$ OPEs; there are no two-loop OPEs. The one-loop $J[1]J[1]$ OPE produces residues that vanish by $\mathfrak{sl}(2,\mathbb{C})_+$ symmetry, since $\spb{2}{2}=0$. The remaining OPE that could produce a pole in eq.~\eqref{jtj2j1j1} is the $J[2]J[1]$ OPE \eqref{j1j2OPE}. This OPE produces several terms. The term which includes $\tilde{K}[1]$ is zero by eq.~\eqref{jtj1kkt1}, while the term which includes $\jt[1]$ vanishes by the results of Sec.~\ref{sec:bcubed1loop}. The remaining terms are non-zero, but their sum vanishes after normally ordering the pairs $J[1]\jt$, $K[1]\tilde{K}$, and $\chi[1]\chi$ in the correlators \eqref{jtjtj1j1}, \eqref{jtj1k1kt}, and \eqref{jtj1chi1chi}. We conclude that eq.~\eqref{jtj2j1j1} holds.

At last, we turn to the vanishing of eq.~\eqref{jt1j1j1j1}. The correlator in eq.~\eqref{jt1j1j1j1} is best viewed as a function of $z_4$, wherein we see that it cannot have singular tree-level OPEs as they result in vanishing three-point two-loop form factors. It also cannot have one-loop OPEs between two $J[1]$'s, as the residue vanishes by Sec.~\ref{sec:bcubed1loop} and eqs.~\eqref{jt1j1kkt} and \eqref{jt1j1chichi}. The one-loop OPE between $\jt[1]$ and $J[1]$ also has a vanishing residue, since 
\begin{equation}
	\la\bthree|:\jt\jt:(z_1)J_{a_2}[1](z_2)J_{a_3}[1](z_3)\ra
	= 0
\end{equation}
from eq.~\eqref{jtjtj1j1}. We conclude that
\begin{equation}
	\F_{4;1}=0
	\,.
\end{equation}

The five-point form factor $\F_{5;1}$ is given by a single CA correlator
\begin{equation}
	\label{F51}
	\F_{5;1} = \la\bthree|\jt_{a_1}(z_1)J_{a_2}[1](z_2)J_{a_3}[1](z_3)J_{a_4}[1](z_4)J_{a_5}[1](z_5)\ra
	\,,
\end{equation}
because all others arise from lower point correlators with an additional $J[0]$ insertion; these are zero since the tree-level OPEs with $J[0]$ land on the lower point vanishing correlators we have already computed.

View the correlation function \eqref{F51} as a function of $z_5$. All tree-level OPEs as $z_5\to z_i$ vanish. Only one-loop OPEs exist as $z_5\to z_i$ for $i=2,3,4$. Taking $z_5\to z_4$ and using the one-loop OPE \eqref{jjloopope}, the terms in the residue are given by normally ordering $J\jt$, $K\tilde{K}$, and $\chi\chi$ in the five-point correlators \eqref{jtjtj1j1-npt}, \eqref{jtj1j1kkt}, and \eqref{jtj1j1chichi}, respectively. Summing these terms gives zero. The limits $z_5\to z_3$ and $z_5\to z_2$ are given by simply relabeling $4\leftrightarrow3$ and $4\leftrightarrow2$ for the $z_5\to z_4$ limit. Hence,
\begin{equation}
	\la\bthree|\jt_{a_1}(z_1)J_{a_2}[1](z_2)J_{a_3}[1](z_3)J_{a_4}[1](z_4)J_{a_5}[1](z_5)\ra = 0
	\,.
\end{equation}
We conclude that
\begin{equation}
	\F_{n;1} = 0
\end{equation}
for all $n$, by induction.

\acknowledgments
NP and JC are supported by the DOE Early Career Research Program under award DE-SC0022924. NP and AM are supported by the Simons Foundation as part of the Simons Collaboration on Celestial Holography.

\appendix
\section{OPEs for modes of component fields}\label{sec:appcoeffs}

The OPEs between the conformally soft modes of the component fields, as fixed by twistorial symmetries and covariance under SUSY transformations, are as follows:
\label{app:components}
\begingroup \allowdisplaybreaks
\begin{align}
	\label{jjloopope}
	\begin{split}
		J_a[t](z)J_b[r](0)  \sim \frac{1}{z}&  \overset{\sum k_j = p_m}{\sum_{ m \geq 1 \quad k_j^i \geq 0}}  \hbar^m \bigg[\overset{(m)}{\underset{(t,r)}{a}}[k_1,...,k_{m+1}]^{d_1 \cdot \cdot \cdot d_{m+1}}_{ab}:J_{d_1}[k_1] \prod_{j=2}^{m+1} \tilde{J}_{d_j}[k_j]:
		\\ 
		& -\overset{(m)}{\underset{(t,r)}{b}}[k_1,...,k_{m+1}]^{d_1 \cdot \cdot \cdot d_{m+1}}_{ab}\delta_i^j:K^i_{d_1}[k_1]\tilde K_{jd_2}[k_2] \prod_{j=3}^{m+1} \tilde{J}_{d_j}[k_j]:
		\\
		& +\frac18 \overset{(m)}{\underset{(t,r)}{b}}[k_1,...,k_{m+1}]^{d_1 \cdot \cdot \cdot d_{m+1}}_{ab}\epsilon_{ijkl}:\chi^{ij}_{d_1}[k_1]\chi^{kl}_{d_2}[k_2] \prod_{j=3}^{m+1} \tilde{J}_{d_j}[k_j]:
		\\
		& +\frac12\overset{(m)}{\underset{(t,r)}{c}}[k_1,...,k_{m+1}]^{d_1 \cdot \cdot \cdot d_{m+1}}_{ab}:\chi^{ij}_{d_1}[k_1]\tilde{K}_{id_2}[k_2]\tilde K_{jd_3}[k_3] \prod_{j=4}^{m+1} \tilde{J}_{d_j}[k_j]:
		\\
		& +\frac1{24}\overset{(m)}{\underset{(t,r)}{d}}[k_1,...,k_{m+1}]^{d_1 \cdot \cdot \cdot d_{m+1}}_{ab}\epsilon^{ijkl}
        \\
        &\hspace{1cm}\times:\tilde  K_{id_1}[k_1]\tilde{K}_{jd_2}[k_2]\tilde K_{kd_3}[k_3]\tilde K_{ld_4}[k_4] \prod_{j=5}^{m+1} \tilde{J}_{d_j}[k_j]:\bigg]
	\end{split}
	\\
	\label{kjloopope}
	\begin{split}
		K^i_a[t](z)J_b[r](0)  \sim \frac{1}{z} & \overset{\sum k_j = p_m}{\sum_{ m \geq 1 \quad k_j^i \geq 0}}  \hbar^m \bigg[\overset{(m)}{\underset{(t,r)}{a}}[k_1,...,k_{m+1}]^{d_1 \cdot \cdot \cdot d_{m+1}}_{ab}:K^i_{d_1}[k_1] \prod_{j=2}^{m+1} \tilde{J}_{d_j}[k_j]:
		\\
		&+ \overset{(m)}{\underset{(t,r)}{b}}[k_1,...,k_{m+1}]^{d_1 \cdot \cdot \cdot d_{m+1}}_{ab}:\chi^{ij}_{d_1}[k_1]\tilde K_{jd_2}[k_2] \prod_{j=3}^{m+1} \tilde{J}_{d_j}[k_j]:
		\\
		& +\frac16\overset{(m)}{\underset{(t,r)}{c}}[k_1,...,k_{m+1}]^{d_1 \cdot \cdot \cdot d_{m+1}}_{ab}\epsilon^{ijkl}
        \\
        &\hspace{2cm}\times:\tilde K_{jd_1}[k_1]\tilde K_{kd_2}[k_1]\tilde K_{ld_3}[k_3] \prod_{j=4}^{m+1} \tilde{J}_{d_j}[k_j]:\bigg]
	\end{split}
	\\
	\label{chijloopope}
	\begin{split}
		\chi^{ij}_a[t](z)J_b[r](0)  \sim \frac{1}{z} & \overset{\sum k_j = p_m}{\sum_{ m \geq 1 \quad k_j^i \geq 0}}  \hbar^m \bigg[\overset{(m)}{\underset{(t,r)}{a}}[k_1,...,k_{m+1}]^{d_1 \cdot \cdot \cdot d_{m+1}}_{ab}:\chi^{ij}_{d_1}[k_1] \prod_{j=2}^{m+1} \tilde{J}_{d_j}[k_j]:
		\\
		& +\frac12\overset{(m)}{\underset{(t,r)}{b}}[k_1,...,k_{m+1}]^{d_1 \cdot \cdot \cdot d_{m+1}}_{ab}\epsilon^{ijkl}:\tilde K_{kd_1}[k_1]\tilde K_{ld_2}[k_2] \prod_{j=3}^{m+1} \tilde{J}_{d_j}[k_j]:\bigg]
	\end{split}
	\\
	\label{ktjloopope}
	\begin{split}
		\tilde K_{ia}[t](z)J_b[r](0)  \sim \frac{1}{z} & \overset{\sum k_j = p_m}{\sum_{ m \geq 1 \quad k_j^i \geq 0}}  \hbar^m \overset{(m)}{\underset{(t,r)}{a}}[k_1,...,k_{m+1}]^{d_1 \cdot \cdot \cdot d_{m+1}}_{ab}:\tilde K_{id_1}[k_1] \prod_{j=2}^{m+1} \tilde{J}_{d_j}[k_j]:
	\end{split}
	\\
	\label{jtjloopope}
	\begin{split}
		\tilde J_{a}[t](z)J_b[r](0)  \sim \frac{1}{z} & \overset{\sum k_j = p_m}{\sum_{ m \geq 1 \quad k_j^i \geq 0}} \hbar^m \overset{(m)}{\underset{(t,r)}{f}}[k_1,...,k_{m+1}]^{d_1 \cdot \cdot \cdot d_{m+1}}_{ab}:\prod_{j=1}^{m+1} \tilde{J}_{d_j}[k_j]:
	\end{split}
	\\
	\label{kkloopope}
	\begin{split}
		K^i_a[t](z)K^j_b[r](0)  \sim -\frac{1}{z} & \overset{\sum k_j = p_m}{\sum_{ m \geq 1 \quad k_j^i \geq 0}}  \hbar^m \bigg[\overset{(m)}{\underset{(t,r)}{a}}[k_1,...,k_{m+1}]^{d_1 \cdot \cdot \cdot d_{m+1}}_{ab}:\chi^{ij}_{d_1}[k_1] \prod_{j=2}^{m+1} \tilde{J}_{d_j}[k_j]:
		\\
		& +\frac12\overset{(m)}{\underset{(t,r)}{b}}[k_1,...,k_{m+1}]^{d_1 \cdot \cdot \cdot d_{m+1}}_{ab}\epsilon^{ijkl}:\tilde K_{kd_1}[k_1]\tilde K_{ld_2}[k_2] \prod_{j=3}^{m+1} \tilde{J}_{d_j}[k_j]:\bigg]
	\end{split}
	\\
	\label{chikloopope}
	\begin{split}
		\chi^{ij}_{a}[t](z)K^k_b[r](0)  \sim \frac{1}{z} & \overset{\sum k_j = p_m}{\sum_{ m \geq 1 \quad k_j^i \geq 0}} \hbar^m \overset{(m)}{\underset{(t,r)}{a}}[k_1,...,k_{m+1}]^{d_1 \cdot \cdot \cdot d_{m+1}}_{ab}\epsilon^{ijkl}:\tilde K_{id_1}[k_1] \prod_{j=2}^{m+1} \tilde{J}_{d_j}[k_j]: 
	\end{split}
	\\
	\label{ktkloopope}
	\begin{split}
		\tilde K_{ia}[t](z)K^j_b[r](0)  \sim -\frac{1}{z} & \overset{\sum k_j = p_m}{\sum_{ m \geq 1 \quad k_j^i \geq 0}}  \hbar^m \overset{(m)}{\underset{(t,r)}{f}}[k_1,...,k_{m+1}]^{d_1 \cdot \cdot \cdot d_{m+1}}_{ab}\delta_i^j:\prod_{j=1}^{m+1} \tilde{J}_{d_j}[k_j]:
	\end{split}
	\\
	\label{chichiloopope}
	\begin{split}
		\chi^{ij}_a[t](z)\chi^{kl}_b[r](0)  \sim \frac{1}{z} & \overset{\sum k_j = p_m}{\sum_{ m \geq 1 \quad k_j^i \geq 0}}  \hbar^m \overset{(m)}{\underset{(t,r)}{f}}[k_1,...,k_{m+1}]^{d_1 \cdot \cdot \cdot d_{m+1}}_{ab}\epsilon^{ijkl}
        \times:\prod_{j=1}^{m+1} \tilde{J}_{d_j}[k_j]:
	\end{split}
\end{align} \endgroup
Recall that
\begin{equation}
	p_m := t+r-(m,m)
	\,,
\end{equation}
where $t=(t^1,t^2)$ and $r=(r^1,r^2)$.

Note that this form of the OPEs is equivalent to that in eq.~\eqref{eq:superOPE}; here we choose to fold the sum over nontrivial permutations of the kinematic variables $k_i$ into the definition of our OPE coefficients, rather than permuting the operators in the normally ordered product. The master coefficient $f$ appearing in the above component OPEs is the same coefficient appearing in eq.~\eqref{eq:superOPE}. 

\subsection {OPE coefficients} 

By employing~\eqref{eq:defect_supercurrent} and~\eqref{eq: Qaction}, we can express the OPE coefficients appearing above in terms of the super-OPE coefficient $f$ seen in eq.~\eqref{eq:superOPE} and defined recursively in eqs.~\eqref{frecursion} and \eqref{f1loop}. The OPE coefficients are\begingroup\allowdisplaybreaks
\begin{align}
	& \overset{(m)}{\underset{(t,r)}{a}}[k_1,...,k_{m+1}]^{d_1 \cdot \cdot \cdot d_{m+1}}_{ab}=\sum_{\pi\in \mathrm{Sh}(1,m)}
	\overset{(m)}{\underset{(t,r)}{f}}
	\bigl[k_{\pi(1)},\ldots,k_{\pi(m+1)}\bigr]^{d_{\pi(1)}\cdots d_{\pi(m+1)}}_{ab}
	\\
	& \overset{(m)}{\underset{(t,r)}{b}}[k_1,...,k_{m+1}]^{d_1 \cdot \cdot \cdot d_{m+1}}_{ab}=\sum_{\pi\in \mathrm{Sh}(2,m-1)}
	\overset{(m)}{\underset{(t,r)}{f}}
	\bigl[k_{\pi(1)},\ldots,k_{\pi(m+1)}\bigr]^{d_{\pi(1)}\cdots d_{\pi(m+1)}}_{ab}
	\\
	& \overset{(m)}{\underset{(t,r)}{c}}[k_1,...,k_{m+1}]^{d_1 \cdot \cdot \cdot d_{m+1}}_{ab}=\sum_{\pi\in \mathrm{Sh}(3,m-2)}
	\overset{(m)}{\underset{(t,r)}{f}}
	\bigl[k_{\pi(1)},\ldots,k_{\pi(m+1)}\bigr]^{d_{\pi(1)}\cdots d_{\pi(m+1)}}_{ab}
	\\
	& \overset{(m)}{\underset{(t,r)}{d}}[k_1,\ldots,k_{m+1}]^{d_1\cdots d_{m+1}}_{ab}
	=\sum_{\pi\in \mathrm{Sh}(4,m-3)}
	\overset{(m)}{\underset{(t,r)}{f}}
	\bigl[k_{\pi(1)},\ldots,k_{\pi(m+1)}\bigr]^{d_{\pi(1)}\cdots d_{\pi(m+1)}}_{ab}
	\,,
\end{align}
where 
\begin{equation}
	\underset{(t,r)}{\overset{(m)}{f}}[k_1,...,k_{m+1}]_{ab}^{d_1 \cdot \cdot \cdot d_{m+1}} = \underset{(t,r)}{\overset{(m)}{f}}(k_1,...,k_{m+1}) K^{d_1 \cdot \cdot \cdot d_{m+1}}_{ab}
	\,.
\end{equation}
We have defined
\begin{align}
	\begin{split}
		\mathrm{Sh}(r,m+1-r) :=&~\{\sig\in S_{m+1}|\sig^{-1}(r+1)<\sig^{-1}(r+2)<\dots<\sig^{-1}(m+1)\}
		\\
		=&~\bigcup_{\sig\in S_r}\sig\shuffle(r+1,r+2,\dotsc,m+1)
		\,,
	\end{split}
\end{align}\endgroup
the set of all permutations of $1,2,\dotsc,m+1$ that preserve the ordering of $r+1,r+2,\dotsc,m+1$.

\section{Recursion Relation for $\overset{(m)}{f}$}
\label{app:recursion}
The simple recursion relation for the master coefficient $\overset{(m)}{f}$ is as follows (see \cite{FP} for more details):\begingroup\allowdisplaybreaks
\begin{align}\label{frecursion}
	&\underset{(r^1,r^2)(t^1,t^2)}{\overset{(m)}{f}}\big(k_1,...,k_{m+1}\big) = \\ \nonumber&-\sum_{j=1}^{t^1}\underset{(r^1,r^2)(t^1-j,t^2)}{\overset{(m-1)}{f}}\big(k_{1},...,k_{m-1},l\big) \underset{(l^1,l^2)(1,0)}{\overset{(1)}{f}}\big(k_{m},\small(k^1_{m+1}+1-j,k^2_{m+1}\small)\big)\\ \nonumber 
	&+\sum_{j=1}^{t^1}\underset{(r^1,r^2)(l^1,l^2)}{\overset{(m-1)}{f}}\big(k_1,...,k_{m}\big) \underset{(t^1-j,t^2)(1,0)}{\overset{(1)}{f}}\big(l,\small(k^1_{m+1}+1-j,k^2_{m+1}\small)\big)\\  \nonumber
	&-\sum_{j=1}^{t^2}\underset{(r^1,r^2)(0,t^2-j)}{\overset{(m-1)}{f}}\big(k_{1},...,k_{m-1},l\big) \underset{(l^1,l^2)(0,1)}{\overset{(1)}{f}}\big(k_{m},\small(k^1_{m+1}-t^1,k^2_{m+1}+1-j\small)\big)
	\,,
\end{align}
where $l$ satisfies $\underset{j = n+1}{\overset{m+1}{\sum}} k_j = l + s - (m-n)$. 
We define the following function:
\begin{equation}
	u(r;k,l) =  \theta(r^i) \theta(k^i) \theta(l^i) \bigg(\frac{1}{16 \pi^2}\bigg) \bigg( \frac{r^1!r^2!(1+k^1+k^2)!(l^1+l^2)!}{k^1!k^2!l^1!l^2!(1+r^1+r^2)!} \bigg) 
	\,.
\end{equation}
In terms of $u$, the one-loop coefficient $\overset{(1)}{\underset{(r,t)}{f}}(k,l)$ is given by:
\begin{multline} 
	\label{f1loop}
	\overset{(1)}{\underset{(r,t)}{f}}(k,l) =  \theta(r^i) \theta(t^i) \theta(k^i) \theta(l^i) \delta^2(r+t-1-k-l) 
	\\
	\times\bigg\{ \sum_{j=1}^{t^1} \bigg(u(r^1+t^1-j,r^2+t^2; k^1,k^2, l^1+1-j,l^2) 
	\\
	- u(t^1-j,t^2;k^1-r^1,k^2-r^2,l^1+1-j,l^2) \bigg) 
	\\
	- \sum_{j=1}^{t^2} u(r^1, r^2+t^2-j; k^1, k^2, l^1-t^1,l^2+1-j ) \bigg\} 
	\,.
\end{multline}\endgroup

\section{Proof of eqs.~\eqref{SWI1minus} \& \eqref{SWIallplus}}
\label{app:proof}

We begin by defining a useful grading for form factors in SDSYM. We recall our superfields and supercurrents in eqs.~\eqref{eq:super-connection-expansion}, \eqref{eq:defect_supercurrent}: 
\begin{equation}
	\cA(Z,\eta)
	\;=\;
	a
	\;+\; \eta_i \lambda^i
	\;+\; \frac{1}{2} \eta_i\eta_j \phi^{ij}
	\;+\; \frac{1}{3!}\epsilon^{ijkl}\eta_i\eta_j\eta_k \tilde\lambda_{l}
	\;+\; \frac{1}{4!}\epsilon^{ijkl}\eta_i\eta_j\eta_k\eta_l\, b,
\end{equation}
\begin{equation}
	\cJ_a(z,\theta)
	=
	J_a(z)
	+\theta_i K_{a}^i(z)
	+\frac{1}{2} \theta_i\theta_j \chi^{ij}_a(z)
	+\frac{1}{3!}\epsilon^{ijkl}\theta_i\theta_j\theta_k \widetilde K_{al}(z)
	+\frac{1}{4!}\epsilon^{ijkl}\theta_i\theta_j\theta_k\theta_l\,\tilde J_a(z).
\end{equation}
These suggest a grading by Grassmann degree $\text{Gr}$, whereby $\text{Gr}(a)=\text{Gr}(J)=0$, $\text{Gr}(\lambda)=\text{Gr}(K)=1, \ldots \text{Gr}(b)=\text{Gr}(\tilde{J})=4$, cf. \cite{WittenTwistor, Boels:2006ir}. 

This grading is exact in SDSYM: referencing our action \eqref{eq:sdN4}, we see the $\text{Gr}$-grading of each term is 4. Note that the action of the chiral supercharge $Q$ lowers the grading by one.

We may consider the Gr-grading of any connected Feynman diagram contributing to a form factor
\begin{equation}\label{eq:allplusapp}
	\langle \mathcal{O} | \mathcal{C}_1[\tilde{\lambda}_1](z_1) \ldots \mathcal{C}_n[\tilde{\lambda}_n](z_n) \rangle
\end{equation}
as follows. If we denote by $V$ the number of bulk vertices arising from the action \eqref{eq:sdN4}, $I$ the number of internal propagators, Gr$_{ext}$ the total charge of the external states $\mathcal{C}_i$ and Gr$(\cO)$ the charge of the operator $\cO$, we have
Gr$_{tot} = \text{Gr}(\cO) + 4V = \text{Gr}_{ext} + 4I$. Defining also the loop number $L$ for a form factor with a single operator insertion as $L = I-V$, we get the relation $\text{Gr}(\cO) - \textrm{Gr}_{ext} = 4 (I-V) = 4 L$. We will use this constraint on Feynman graph topology in what follows. For the remainder of this section, we specialize to $\cO_m := \text{tr}(B^m)$\footnote{We neglect the overall normalization constants as they will not affect our argument.}, so $\text{Gr}(\cO_m) = 4 m$ and $\text{Gr}(Q\cdot\cO_m)= 4m-1$. 

To establish our claim \eqref{SWIallplus}, consider the form factor
\begin{equation}
	\langle \cO_m| K_1 J_2 \ldots J_n \rangle
\end{equation} with $\text{Gr}_{ext} = 1$. Assuming the vacuum preserves supersymmetry, we write\footnote{We will use the notation of graded brackets so that the appropriate commutator or anticommutator with $Q$, depending on the parity of the state, is understood.}
\begin{align}
	0 &= \langle [Q_{I \alpha}, \cO_m K_1 J_2 \ldots J_2] \rangle \\ \nonumber
	&=\langle (Q\cdot\cO_m)| K_1 J_2 \ldots J_n \rangle + \langle \cO_m | [Q_{I \alpha}, K_1^I]J_2 \ldots J_n \rangle \\ \nonumber
	&= \langle (Q\cdot\cO_m)| K J_2 \ldots J_n \rangle + \lambda_{1 \alpha}\langle \cO_m | J_1 \ldots J_n \rangle
\end{align} where we have used the fact that the chiral supercharge annihilates all positive-helicity gluons, and acts as $[Q_{I \alpha}, K^J_r] = \delta^J_I \lambda_{r \alpha}J_{r}$.

The first term is a form factor with an insertion of the descendant operator $Q\cdot \cO_m$. Referring back to our grading, this descendant form factor must satisfy $1 = (4m-1)- 4L \rightarrow 4L = 4m-2$, but there is no integer loop number for which this constraint can be satisfied. Therefore, the descendant form factor vanishes and we are left with equation \eqref{SWIallplus}. 

We use a similar argument to establish eq.~\eqref{SWI1minus}. Let us now consider the form factor
\begin{equation}\label{eq:oneminusapp}
	\langle \cO_m| \tilde{J}_1K^I_2 J_3 \ldots J_n \rangle. 
\end{equation} We again apply the chiral supercharge to obtain
\begin{align}
\begin{split}
    0 &= \langle Q\cdot\cO_m | \tilde{J}_1K^I_2 J_3 \ldots J_n \rangle + \langle \cO_m| [Q_{I \alpha}, \tilde{J}_1] K_2^I J_3 \ldots J_n \rangle + \langle \cO_m| \tilde{J}_1 [Q_{I \alpha}, K_2^{I}]J_3 \ldots J_n \rangle \\
	&= \langle Q\cdot\cO_m | \tilde{J}_1K^I_2 J_3 \ldots J_n \rangle + \lambda_{1 \alpha}\langle \cO_m| \tilde{K}_{I, 1} K_2^I J_3 \ldots J_n \rangle + \lambda_{2\alpha}\langle \cO_m | \tilde{J}_1 J_2 \ldots J_n \rangle.
\end{split}
\end{align}
The descendant form factor again vanishes since the Grassmann charge constraint implies $4L = 4 m -6$, which cannot be satisfied for integer loop number. We are left with a two-term spinor equation. Contracting the equation in turn with $\epsilon^{\alpha \beta}\lambda_{2 \beta}$ forces the two-fermion form factor to be zero, and contracting $\epsilon^{\alpha \beta}\lambda_{1 \beta}$ requires the one-minus form factor to vanish as well, establishing eq.~\eqref{SWI1minus}. 

We remark that for higher numbers of negative-helicity gluons, applying this logic will result in the equation that a sum of three or more form factors, each weighted by different spinors, equals zero. There are not enough constraints to force all the coefficients to vanish; instead, the SWI can then be used to derive relations among the terms.

Finally, this argument admits a simple generalization to the case when the operator is $\prod_{i=1}^k\text{tr}(B^{m_i})(x_i)$ for non-coincident operator insertions. Defining $N_B = \sum_{i=1}^k m_i$, the selection rule becomes $4 N_B + 4(V-I) = \text{Gr}_{ext}$. Applying the SWI to form factors with the same external states as eq.~\eqref{eq:allplusapp} (respectively eq.~\eqref{eq:oneminusapp}) produces a sum of descendant form factors each with gradings $4 N_B -2 = 4(I-V)$ (resp. $4N_B-6 = 4(I-V)$), so they each vanish. The remainder of the proof proceeds identically.

% Bibliography

\bibliographystyle{JHEP}
\bibliography{bootstrap.bib}

\end{document}